\documentclass[twocolumn]{aastex63}
\usepackage{xcolor}
\usepackage{makecell}
\usepackage{pbox}

\received{\today}
\submitjournal{ApJ}
\shorttitle{ASASSN-14ko}
\shortauthors{Payne et al.}
\graphicspath{{./}{figures/}}

\def\gal{ESO 253$-$G003}

\begin{document}

\title{Chandra, HST/STIS, NICER, Swift, and TESS Detail the Flare Evolution of the Repeating Nuclear Transient ASASSN-14ko}

\correspondingauthor{Anna V. Payne}
\email{avpayne@hawaii.edu}

\author[0000-0003-3490-3243]{Anna V. Payne}
\altaffiliation{NASA Graduate Fellow}
\affiliation{Institute for Astronomy, University of Hawai\`{}i at Manoa, 2680 Woodlawn Dr., Honolulu, HI 96822}

\author[0000-0002-4449-9152]{Katie~Auchettl}
\affiliation{School of Physics, The University of Melbourne, Parkville, VIC 3010, Australia}
\affiliation{ARC Centre of Excellence for All Sky Astrophysics in 3 Dimensions (ASTRO 3D)}
\affiliation{Department of Astronomy and Astrophysics, University of California, Santa Cruz, CA 95064, USA}

\author[0000-0003-4631-1149]{Benjamin J. Shappee}
\affiliation{Institute for Astronomy, University of Hawai\`{}i at Manoa, 2680 Woodlawn Dr., Honolulu, HI 96822}

\author[0000-0001-6017-2961]{Christopher~S.~Kochanek}
\affiliation{Department of Astronomy, The Ohio State University, 140 West 18th Avenue, Columbus, OH 43210, USA}
\affiliation{Center for Cosmology and AstroParticle Physics, The Ohio State University, 191 W.\ Woodruff Ave., Columbus, OH 43210, USA}

\author{Patricia T. Boyd}
\affiliation{Laboratory for Exoplanets and Stellar Astrophysics, NASA Goddard Space Flight Center, Greenbelt, MD 20771, USA}

\author[0000-0001-9206-3460]{Thomas~W.-S.~Holoien}
\altaffiliation{NHFP Einstein Fellow}
\affiliation{The Observatories of the Carnegie Institution for Science, 813 Santa Barbara St., Pasadena, CA 91101, USA}

\author[0000-0002-9113-7162]{Michael M. Fausnaugh} 
\affiliation{Department of Physics, and Kavli Institute for Astrophysics and Space Research, Massachusetts Institute of Technology, Cambridge,
MA 02139, USA}

\author[0000-0002-5221-7557]{Chris Ashall}
\affiliation{Institute for Astronomy, University of Hawai\`{}i at Manoa, 2680 Woodlawn Dr., Honolulu, HI 96822}

\author[0000-0001-9668-2920]{Jason T. Hinkle}
\affiliation{Institute for Astronomy, University of Hawai\`{}i at Manoa, 2680 Woodlawn Dr., Honolulu, HI 96822}

\author[0000-0001-5661-7155]{Patrick J. Vallely}
\affiliation{Department of Astronomy, The Ohio State University, 140 West 18th Avenue, Columbus, OH 43210, USA}

\author{K.~Z.~Stanek}
\affiliation{Department of Astronomy, The Ohio State University, 140 West 18th Avenue, Columbus, OH 43210, USA}
\affiliation{Center for Cosmology and AstroParticle Physics, The Ohio State University, 191 W.\ Woodruff Ave., Columbus, OH 43210, USA}

\author[0000-0003-2377-9574]{Todd~A.~Thompson}
\affiliation{Department of Astronomy, The Ohio State University, 140 West 18th Avenue, Columbus, OH 43210, USA}
\affiliation{Center for Cosmology and AstroParticle Physics, The Ohio State University, 191 W.\ Woodruff Ave., Columbus, OH 43210, USA}

\begin{abstract}
    ASASSN-14ko is a nuclear transient at the center of the AGN \gal{} that undergoes periodic flares. Optical flares were first observed in 2014 by the All-Sky Automated Survey for Supernovae (ASAS-SN) and their peak times are well-modeled with a period of $115.2^{+1.3}_{-1.2}$ days and period derivative of $-0.0026 \pm 0.0006$. Here we present ASAS-SN, \textit{Chandra}, \textit{HST}/STIS, \textit{NICER}, \textit{Swift}, and \textit{TESS} data for the flares that occurred in December 2020, April 2021, July 2021, and November 2021. The \textit{HST}/STIS UV spectra evolve from blue shifted broad absorption features to red shifted broad emission features over $\sim$10 days. The \textit{Swift} UV/optical light curves peaked as predicted by the timing model, but the peak UV luminosities varied between flares and the UV flux in July 2021 was roughly half the brightness of all other peaks. The X-ray luminosities consistently decreased and the spectra became harder during the UV/optical rise but apparently without changes in absorption. Finally, two high-cadence \textit{TESS} light curves from December 2020 and November 2018 showed that the slopes during the rising and declining phases changed over time, which indicates some stochasticity in the flare's driving mechanism. ASASSN-14ko remains observationally consistent with a repeating partial tidal disruption event, but, these rich multi-wavelength data are in need of a detailed theoretical model.  
\end{abstract}

\section{Introduction}

A tidal disruption event (TDE) occurs when a star crosses the tidal radius of a central supermassive black hole (SMBH) and is torn apart by the SMBH's tidal forces \citep{hills75, rees88, phinney89, evans89}. This results in roughly half of the stellar mass becoming unbound and the remainder forming an accretion disk that feeds the SMBH, producing a luminous multi-wavelength flare. TDEs exhibit diverse behaviors across the electromagnetic spectrum. This includes a wide range of X-ray brightening timescales and luminosities (e.g., \citealt{auchettl2017, holoien2018, wevers2019, kajava2020, hinkle2021_29dj}); optical spectral properties such as the presence or absence of Hydrogen, Helium, and Bowen fluorescence lines (e.g., \citealt{gezari12, arcavi14, holoien14b, holoien16, holoien16b,  brown16, brown17, holoien2018, leloudas2019, nicholl2019, holoien2020, vanvelzen2021, hinkle2022atlas18mlw}); and differences in their blackbody luminosity, radius, and temperature evolution \citep{hinkle20a, hinkle2021}. 

The ultraviolet (UV) spectral properties of TDEs are less well explored. The small sample of TDEs with UV spectra includes ASASSN-14li \citep{cenko2016}, iPTF-16fnl \citep{brown2018}, iPTF-15af \citep{blagorodnova2019},  PS18kh (AT 2018zr: \citealt{hung2019}), and ZTF19abzrhgq (AT 2019qiz: \citealt{hung2020}), in addition to the ambiguous nuclear transient (ANT) ASASSN-18jd which showed properties indicative of both TDEs and active galactic nuclei (AGNs) (AT 2018bcb: \citealt{neustadt2020}). UV spectra of TDEs are generally characterized by a hot continuum and broad lines including Ly$\alpha$, \ion{N}{5} $\lambda1240$, \ion{Si}{4} $\lambda1400$, and \ion{C}{4} $\lambda1550$ in emission and absorption. Somewhat surprisingly, they show little temporal evolution when multiple epochs are available. To date, UV spectra have only been obtained after maximum light and during the decline, weeks to months after maximum light. 

UV spectra of AGN are characterized by broad and narrow emission lines on top of a roughly power-law continuum spectrum (e.g., \citealt{krolik1999}). Typical spectral features include Ly$\alpha$ $\lambda 1215$, \ion{C}{4} $\lambda 1549$, \ion{Si}{4} $\lambda 1400$, \ion{N}{5} $\lambda 1240$, \ion{O}{5} $\lambda 1035$, \ion{C}{3}] $\lambda 1908$, and \ion{Mg}{2} $\lambda 2800$, as well as broad absorption lines generally on the blue wing of the lines. Both the lines and the continuum are time variable.  The vast majority show modest levels of continuum variability that are reasonably, but not perfectly, modeled by a damped random walk stochastic process (e.g., \citealt{kelly08, kozlowski10, macleod10, zu13}).  The UV continuum variability in turn drives changes in the broad emission lines which are frequently studied using reverberation mapping (\citealt{blandford82, peterson93, peterson04}). In particular, the emission lines are observed to narrow as the AGN becomes brighter, as expected from photoionization models. 

A small fraction of AGNs are ``changing look quasars'' where the structure of the emission lines completely changes between narrow and broad line dominated spectra, usually with a significant change in the continuum brightness (e.g., \citealt{bianchi2005, denney2014, shappee14, macleod2016, hon2022}).  In the case of ``rapid turn-on'' events, the blue continuum and broad lines appear on time scales of a few months (e.g., \citealt{gezari2017, frederick2019, gromadzki2019,  trakhtenbrot2019nature, ricci2020}).  There are also ANTs which share characteristics of both TDEs and more normal AGN variability (\citealt{neustadt2020,hinkle20hx,holoien2021,yu2022}).

Here we present observations of the latest four flares of ASASSN-14ko, a nuclear transient at the center of the AGN \gal{} that undergoes periodic flares \citep{payne2021a, payne2021b}. This transient was initially discovered by the All-Sky Automated Survey for Supernovae (ASAS-SN, \citealt{shappee14, kochanek17}) in 2014 but classified as a Type IIn supernova with a blue continuum projected very close to the nucleus of a Type 2 Seyfert, although AGN activity was not ruled out as a possibility \citep{holoien14ATELc}. However, the subsequent seven years of ASAS-SN data revealed that the flare was not a one-time event -- the flares recur at regular intervals well-fit by a timing model with a mean period of roughly $115$ days and a negative period derivative. Each flare is consistently characterized by a single UV/optical brightening event that rapidly rises and smoothly declines over $\sim40$ days. The X-ray flux dims rapidly (days) during the UV rise and then recovers. The host galaxy \gal{} is a complex merger remnant with two AGNs and a large tidal arm \citep{tucker2020}. The brighter northeastern nucleus is the source of the periodic flares \citep{payne2021b}.  

In this paper, we present the data and analysis of the four flares which peaked in December 2020, April 2021, July 2021, and November 2021. In Section \ref{observations}, we discuss the data used in this paper. In Section \ref{light_curve}, we show the photometric light curves, updated timing model, and how the latest flares compare to previous flares. In Section \ref{uvspectra} we analyze the UV spectra, and in Section \ref{xray} we investigate the X-ray emission. In Section \ref{discussion} we discuss how the latest flares fit into interpretations of ASASSN-14ko's origins and we give our conclusions in Section \ref{conclusion}. For a flat $\Omega_m = 0.3$ universe, the luminosity distance is $\approx 188$ Mpc and the projected scale is $\approx 0.85$ kpc/arcsec. The Galactic extinction is $\mathrm{A}_{\mathrm{V}} = 0.116$ mag \citep{schlafly11}. 

\section{Observations} \label{observations}

The following sections describe the data used in this analysis. All photometric data discussed here are presented in Table \ref{tab:short_phot}.

\subsection{ASAS-SN}

ASAS-SN is an ongoing all-sky survey that uses four 14-cm telescopes on a common mount to monitor the sky to find bright, nearby transients \citep{shappee14, kochanek17}. There are currently five units located in Hawai\`{}i, Chile, Texas, and South Africa that are hosted by the Las Cumbres Observatory global telescope network \citep{brown13}. ASAS-SN images are processed with a fully automatic pipeline that utilizes the ISIS image subtraction package \citep{alard98, alard00}. Reference images are used to subtract the background and host emission from all science images, and aperture photometry is then performed using the IRAF \texttt{apphot} package \citep{tody1986, tody1993}. The data were calibrated using stars from the AAVSO Photometric All-Sky Survey (APASS; \citealt{henden15}). All low-quality images were inspected by eye, and images affected by clouds or other systematic problems are removed.

\subsection{HST/STIS}

We observed ASASSN-14ko using STIS and the FUV/NUV MAMA detectors (HST IDs: 16451, 16498; PI: Shappee). We used the $52.\!\!''0 \times 0.\!\!''2$ slit and the G140L (1425 \AA, FUV-MAMA) and G230L (2376 \AA, NUV-MAMA) gratings. FUV and NUV spectra were obtained on 2020-12-18, 2020-12-22, 2020-12-29, 2021-03-06, and FUV-only spectra were obtained on 2021-03-29 and 2021-04-01. Each visit consisted of 2 or 3 exposures per grating with exposure times ranging from 336 to 482 seconds for the G230L grating and 200 to 1732 seconds for the G140L grating. We used the one-dimensional spectra produced by the HST pipeline since the trace of ASASSN-14ko was clearly present in the two-dimensional frames. For each epoch, we combined the individual exposures with an inverse-variance-weighted average of the one-dimensional spectra and then merged the FUV and NUV channels.  

\subsection{Swift XRT \& UVOT}

We requested \textit{Swift} ToO observations (ToO ID: 14775, 14971, 15393, 15636, 16037, 16546; PI: Payne) using the UltraViolet/Optical Telescope (UVOT) and the X-Ray Telescope (XRT) to coincide with the predicted December 2020, April 2021, July 2021, and November 2021 flares. The UVOT data were obtained in six  filters \citep{poole08}: $V$ (5425.3 \AA), $B$ (4349.6 \AA), $U$ (3467.1 \AA), $UVW1$  (2580.8 \AA), $UVM2$  (2246.4 \AA),  and $UVW2$ (2054.6 \AA). The wavelengths used here are the pivot wavelengths calculated by the SVO Filter Profile Service \citep{rodrigo2012}. The \textit{Swift} team announced in November 2020 that a loss of sensitivity over time requires an updated photometric correction for the three UVOT UV filters, and we used the updated correction for this analysis (see \citealt{hinkle2021}). We used the HEAsoft (\hspace{-1mm}\citealt{heasarc2014}) software task \textit{uvotsource} to extract the source counts using a $16.\!\!''0$ radius aperture and used a sky region of $\sim$ $40.\!\!''0$ radius to estimate and subtract the sky background. 

The XRT data were reduced using the HEAsoft task \textit{xrtpipeline}. We obtained a background subtracted count rate from each observation using a source region with a radius of 50'' centered on the position of ASASSN-14ko and a 150'' radius source free background region centered at ($\alpha$,$\delta$)=($05^{h}25^{m}09.87^{s}, -45^{\circ}56'47.94''$). All count rates were corrected for encircled energy fraction. 

\subsection{TESS}

The Transiting Exoplanet Survey Satellite (\textit{TESS}, \citealt{ricker2014}) observed ASASSN-14ko during Sectors 31-33, which occurred between 2020-10-21 and 2021-01-13, with the flare occurring during Sectors 32-33. Following a similar procedure to that used for the ASAS-SN data, we used the ISIS package to perform image subtraction on the 10-minute cadence full frame images (FFIs) to obtain high fidelity light curves as described in \citet{vallely2019,vallely2021}. This is the second flare captured by \textit{TESS}, following an earlier observation during Sectors 3-5 of the flare that peaked in November 2018 \citep{payne2021b}.  

\begin{table}[t]
\centering
\caption{Photometry of ASASSN-14ko used in this analysis.  The luminosities are in units of $10^{40}$ $\mathrm{ergs}$  $\mathrm{s}^{-1}$ and the X-ray luminosities are between $0.3-10.0$ keV.  Only the first observation in each band is shown here to demonstrate its form and content. Table to be published in its entirety in machine-readable form in the online journal.    }
\begin{tabular*}{\columnwidth}{l l l l}
\toprule

JD  & Band & \pbox{25mm}{L$_{\lambda}$ ($\mathrm{ergs}$  $\mathrm{s}^{-1}$ )}  & \pbox{35mm}{L$_{\lambda}$ Error ($\mathrm{ergs}$  $\mathrm{s}^{-1}$)} \\

\hline
58983.65 & X-ray & 186 & 50.3 \\
58983.65 & $W2$ & \hspace{3.mm}4.89 &  \hspace{1.5mm}0.29 \\
58983.68 &  $M2$ &  \hspace{3.mm}3.99 &  \hspace{1.5mm}0.43 \\
58983.65 &  $W1$ &  \hspace{3.mm}2.29 &  \hspace{1.5mm}0.16 \\
58983.65 &  $U$ &  \hspace{3.mm}1.15 &  \hspace{1.5mm}0.09 \\
58983.65 &  $B$ &  \hspace{3.mm}0.82 &  \hspace{1.5mm}0.08 \\
58983.66 &  $V$ &  \hspace{3.mm}0.62 &  \hspace{1.5mm}0.08 \\
58849.34 &  $g$ &  \hspace{3.mm}0.63 &  \hspace{1.5mm}0.08 \\
58425.46 & $I_{\mathrm{TESS}}$ & \hspace{3.mm}0.019 & \hspace{1.5mm}0.005 \\

\hline 

\end{tabular*}
\label{tab:short_phot}
\end{table}

\subsection{NICER}

ASASSN-14ko was also observed using the  Neutron star Interior Composition ExploreR (\textit{NICER}: \citealt{gendreau2012}) and its X-ray timing instrument (XTI). ASASSN-14ko was observed a total of 115 times between 2020 Dec 10 and 2021 Dec 01 (ObsIDs:  3201740101$-$3201740118,4662010101$-$4662022801, PI: Payne), for a total cumulative exposure of 277.7ks. 

The data were reprocessed using the \textsc{nicerdas} version 8c and the task \textsc{nicerl2}. Here standard filtering criteria were used\footnote{See \citet{bogdanov2019} or Section 2.7 of \citet{hinkle2021_29dj}}, as well as the latest gain and calibrations files. Time averaged spectra and count rates were extracted using \textsc{xselect}. We used the ARF (nixtiaveonaxis20170601v005.arf) and RMF (nixtiref20170601v003.rmf) files that are available with the \textit{NICER} CALDB. All spectra were grouped using a minimum of 20 counts per energy bin. As \textit{NICER} is a non-imaging instrument, background spectra were generated using the background modeling tool \textsc{nibackgen3C50}\footnote{\url{https://heasarc.gsfc.nasa.gov/docs/nicer/tools/nicer_bkg_est_tools.html}}.

\subsection{Chandra}
Finally, we acquired two \textit{Chandra} ACIS-S DDT observations (observation IDs: 24875, 24876; PI: Payne). The first observation occurred on 2020-12-26 with an exposure time of 27.69 ks, and the second observation occurred on 2021-01-21 with an exposure time of 28.68 ks. The data were analyzed using CIAO 4.13 and CALDB 4.9.4. The data were reduced using the CIAO command \textsc{chandra\_repro}, while spectra were extracted from each reprocessed observation using the CIAO command \textsc{specextract}. All spectra were grouped using a minimum of 20 counts per bin.  To analyze the spectral data from \textit{Swift}, \textit{NICER} and \textit{Chandra} data, we used the X-ray spectral fitting package (XSPEC) version 12.12.0 \citep{arnaud1996} and $\chi^{2}$ statistics.

\section{UV/Optical Light Curve Analysis} \label{light_curve}

The December 2020, April 2021, July 2021, and November 2021 flares are the third, fourth, fifth, and sixth events, respectively, in which the UV properties were observed in conjunction with the optical evolution. These data enable us to search for similarities and differences in the UV properties of six flares.

\subsection{UV/Optical Evolution} \label{timingmodel}

\begin{figure}
    \centering
    \includegraphics[width=\columnwidth]{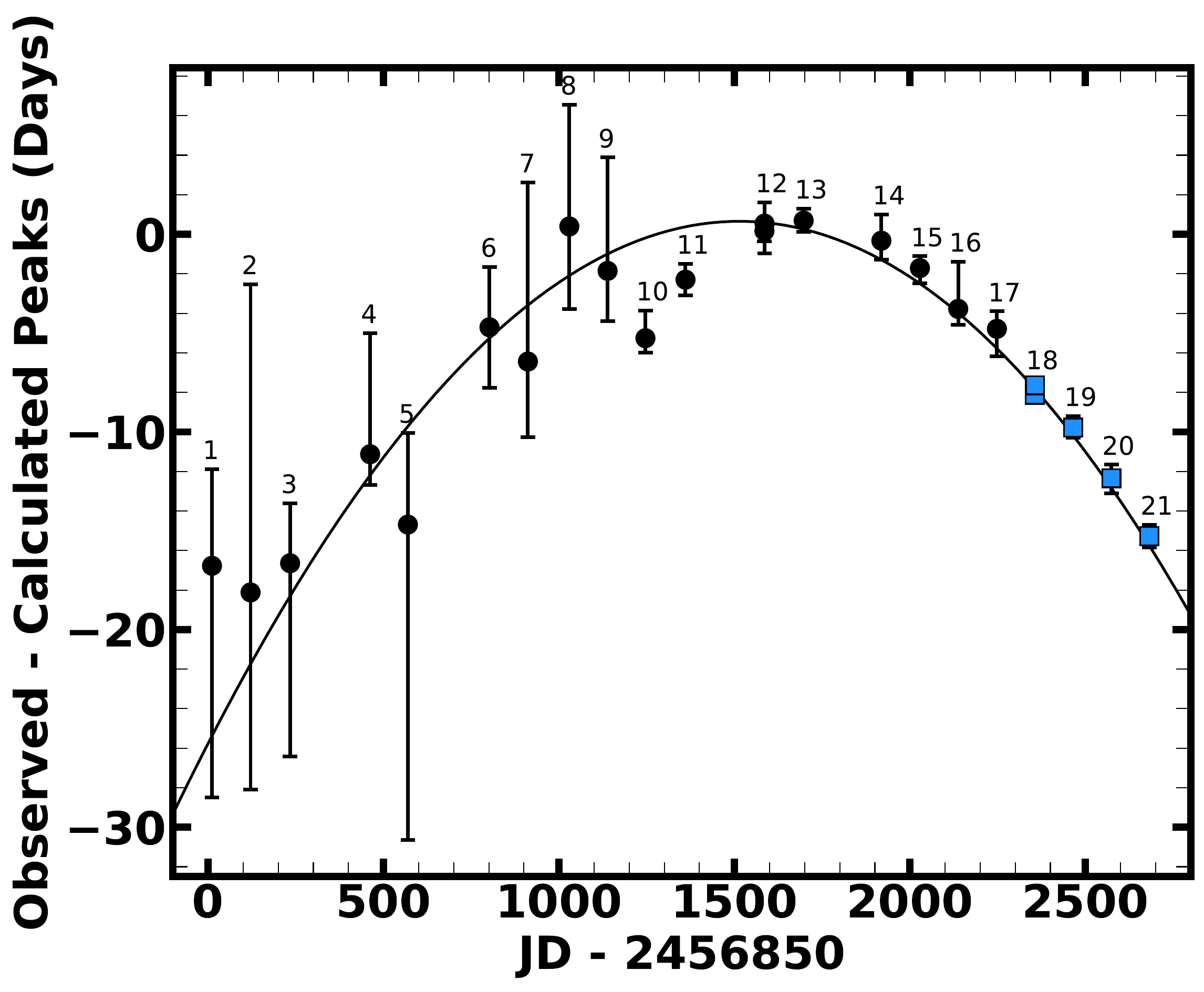}
    \caption{O-C plot for all of ASASSN-14ko's observed outbursts comparing the time of the observed peak with the estimated peak if we assume a constant period. The black circles correspond to the the flares reported in \citet{payne2021a, payne2021b} and the blue squares represent the flares reported here. The solid line is the model described in Section \ref{timingmodel}.  }
    \label{fig:o-c_plot}
\end{figure}

We showed previously in \citet{payne2021a} and \citet{payne2021b} that ASASSN-14ko's flares are essentially periodic, so each new flare is another opportunity to update the timing model of the optical peaks. In \citet{payne2021a} we found significant residuals between the observed and calculated peak times (Observed - Calculated, O-C), which is indicative of a period derivative. To analyze how this trend continued, we measured the optical peaks of the most recent flares that occurred in December 2020, April 2021, July 2021, and November 2021 (as shown on Figure \ref{fig:o-c_plot}: flares 18-21) by fitting the ASAS-SN $g$-band light curves with a fifth-order polynomial and measured the errors on the peak times by bootstrap resampling the data. These peaks occurred on $59205.7 \pm 0.19$, $59315.2^{+0.6}_{-0.5}$,  $59433.3^{+6}_{-3}$, and $59532.2 \pm 0.9$, respectively. The comparatively larger error on July 2021's peak timing is due to a gap in the ASAS-SN light curve due to bad weather, so we also measured this event's peak timing using the \textit{Swift} \textit{B}-band light curve whose peak time was MJD $59423.9^{+0.7}_{-0.8}$. In total, there are now 21 flares observed by ASAS-SN since 2014 after combining the new flares reported here with those in \citet{payne2021a} and \citet{payne2021b}. 

For the flares that occurred during November 2018 (flare 12) and December 2020 (flare 18), \textit{TESS} also observed these flares so the peak times measured from the \textit{TESS} light curves were also included alongside the peak times measured from ASAS-SN. We fit the model 
\begin{equation}
    t = t_0 + nP_0 + \frac{1}{2}n^2 P_0 \dot{P} + \frac{1}{6} n^3 P_0 \dot{P}^2,
\end{equation}
and obtain starting time $t_0 = 2456852^{+10}_{-11}$ JD, mean period $P_0 = 115.2^{+1.3}_{-1.2}$ days, and a period derivative $\dot{P} = -0.0026 \pm 0.0006 $. The errors on the peak times were expanded in quadrature by 0.8 days to obtain a reduced $\chi^2$ of unity for 10 degrees of freedom. The O-C diagram for all observed peaks with the updated timing model is shown in Figure \ref{fig:o-c_plot}. The parabolic trend in the O-C timings continues. This timing model predicts that the next flares will peak in the optical on MJD $59639.7$ (UT 2022-03-01.2), $59747.5$ (UT 2022-06-17.0), $59855.9$ (UT 2022-10-03.4), $59962.1$ (UT 2023-01-17.6), and $60069.0$ (UT 2023-05-04.5). 

In \citet{payne2021b}, we had concluded that $\dot{P}$ had changed and used only the flare timings obtained after flare 9 for the timing model to predict when subsequent flares would occur. With these four new flares it now seems that the timing of flare 10 and flare 11 were anomalous and the older (flares 1-9) and recent flares (flares 12-21) are consistent with a single $\dot{P}$. Certainly for a repeating TDE we should expect some stochasticity in $\dot{P}$ --- it should be problematic for this interpretation if $\dot{P}$ did not vary. 

\begin{figure*}
    \centering
    \includegraphics[width=0.97\linewidth]{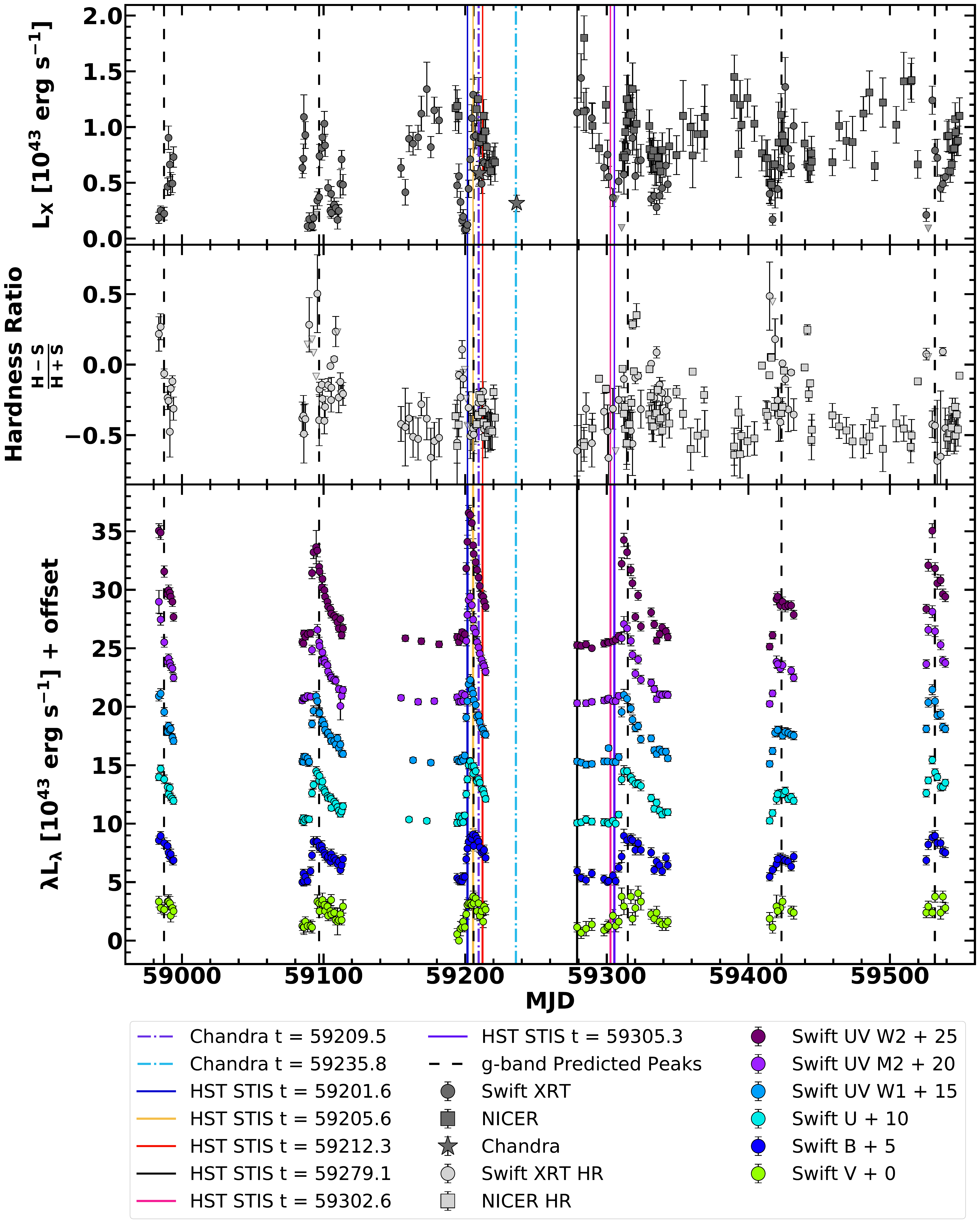}
    \caption{The \textit{Swift} and \textit{NICER} X-ray and UV luminosities and hardness ratios for the six flares in 2020-2021. The vertical dashed lines indicate the  $g$-band peaks predicted from the updated timing model described in Section \ref{light_curve}. The vertical solid bars are color-coded to the \textit{HST}/STIS spectra epochs shown in Figure \ref{fig:all_STIS} and the \textit{Chandra} spectra epochs shown in Figure \ref{fig:chandra_fig}. Upper limits are indicated by downward facing triangles. The X-ray luminosities cover $0.3-10.0$ keV and the hard and soft X-rays are defined between $0.3-2.0$ keV and $2.0-10.0$ keV respectively.  }
    \label{fig:full_LC}
\end{figure*}

\begin{figure*}[hbtp]
    \centering
    \includegraphics[width=0.89\linewidth]{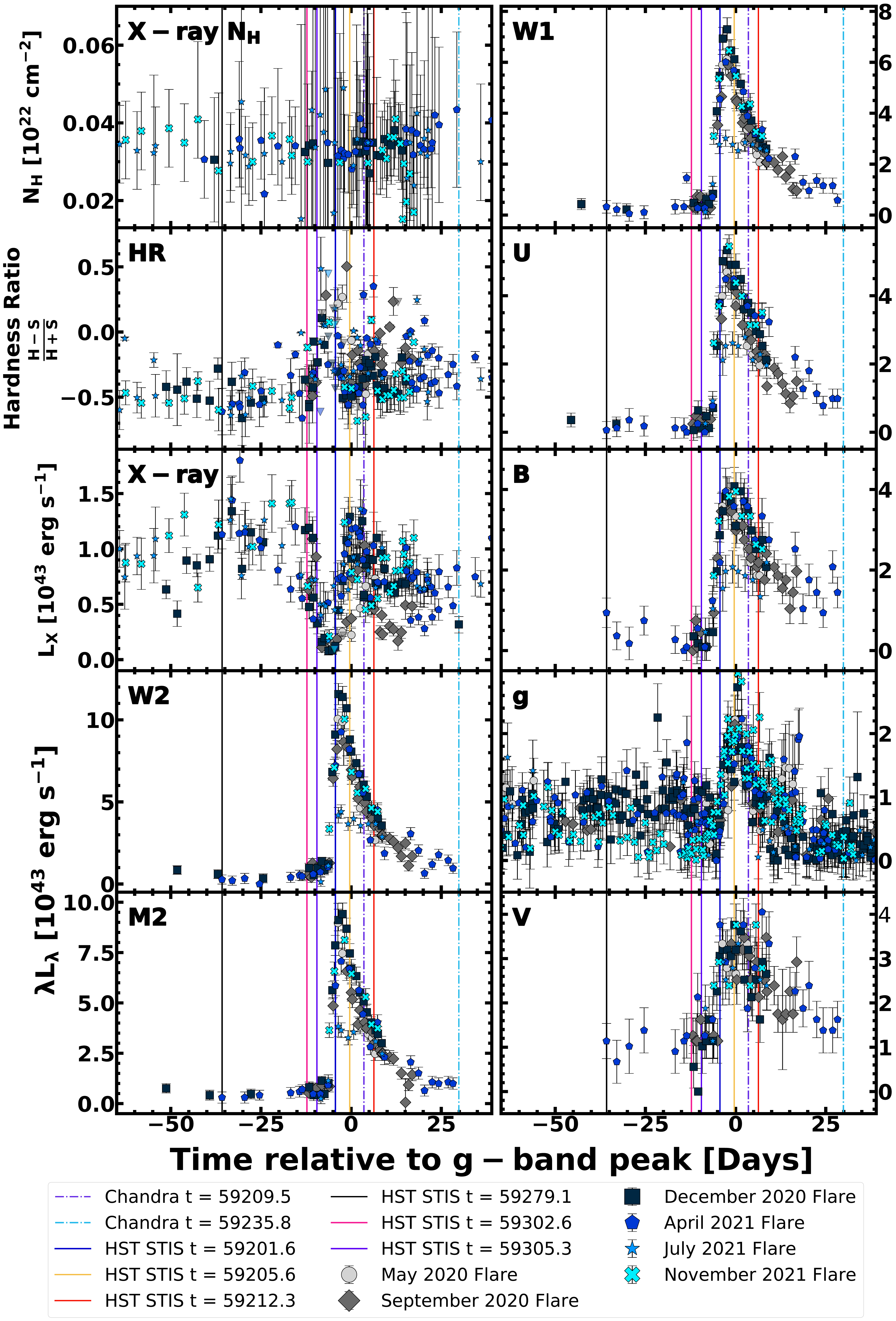}
    \caption{\textit{Swift} XRT and \textit{NICER} X-ray and host-subtracted \textit{Swift} UVOT, and ASAS-SN $g$-band light curves aligned using the updated timing model described in Section \ref{light_curve} including May 2020 (circles), September 2020 (triangles), December 2020 (squares), April 2021 (pentagons), July 2021 (stars), and November 2021 (Xs) flares. The vertical bars are color-coded to the \textit{HST}/STIS spectra epochs shown in Figure \ref{fig:all_STIS} and the \textit{Chandra} spectra epochs shown in Figure \ref{fig:chandra_fig}.    }
    \label{fig:phasefold_phot}
\end{figure*}

After updating the timing model following each new flare, we used the latest predictions to schedule the observations of the next flare. Figure \ref{fig:full_LC} shows the UV/optical host-subtracted light curves for the six most recent flares from November 2021, July 2021, April 2021, December 2020, September 2020, and May 2020. In order to directly compare the flare evolution across all flares, we stacked the host-subtracted light curves using the optical peak times predicted by the timing model as shown in Figure \ref{fig:phasefold_phot}. The flares are consistently characterized by a single, asymmetric UV and optical rise and fall without obvious substructure. The six flares are not identical. The peak UV luminosity changes and the differences are largest in the shortest wavelength $W2$ and $M2$ UV bands. This contrasts with the optical bands, which have not shown discernible changes in peak luminosity between flares. The July 2021 flare was significantly less luminous than all the others across all bands except for the $V$-band, which was noisier. The July 2021 flare was the first indication that the flares can have significantly different peak luminosities. 

\begin{figure}
    \centering
    \includegraphics[width=0.97\linewidth]{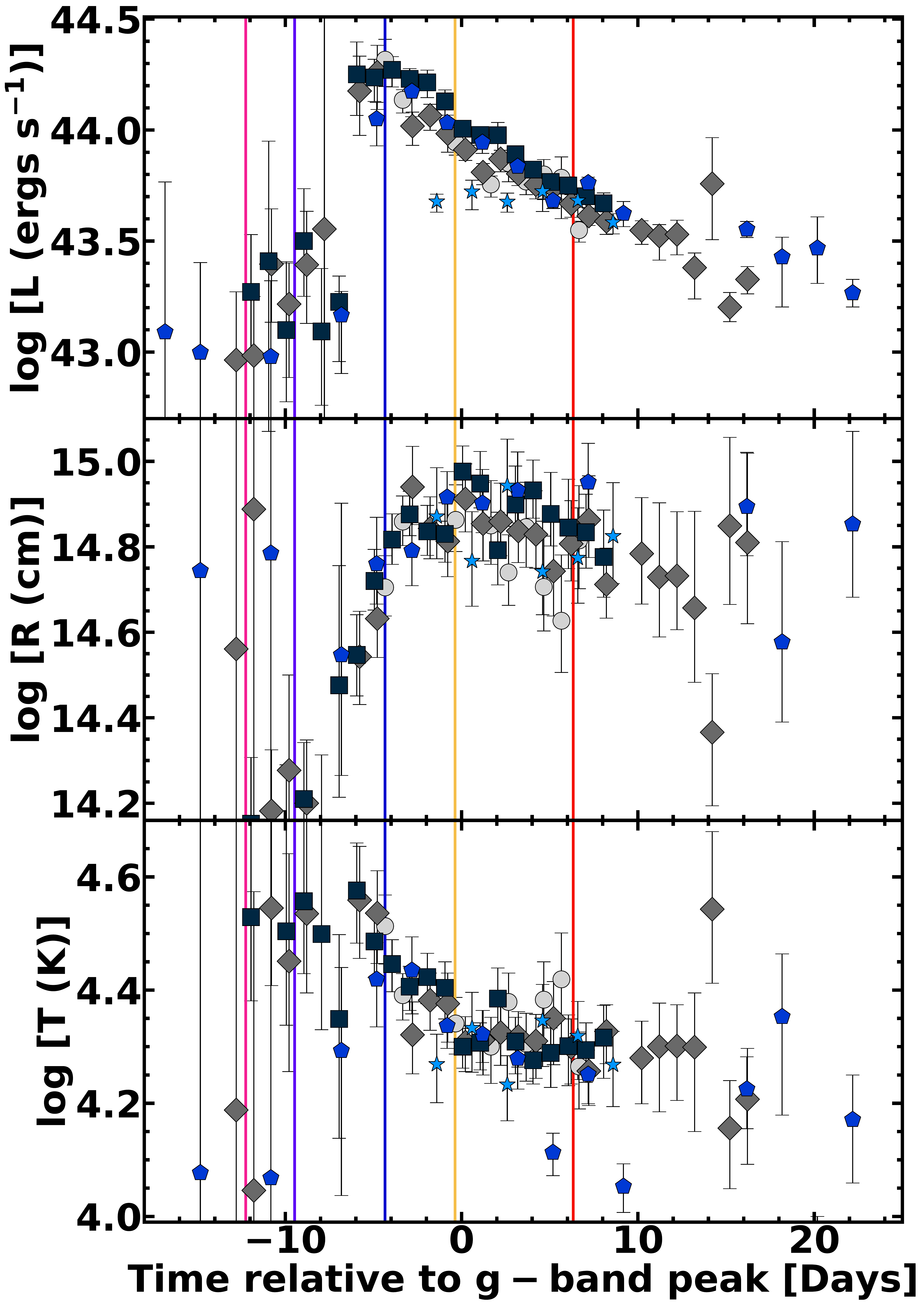}
    \caption{Evolution of ASASSN-14ko's luminosity (\textit{top}), effective radius (\textit{middle}), and temperature (\textit{bottom}) from blackbody fits to the host-subtracted UV/optical \textit{Swift} data. The colors and symbols match those used in Figure \ref{fig:phasefold_phot} and the data are aligned using the timing model described in Section \ref{light_curve}.    }
    \label{fig:bbody_phasefold}
\end{figure}

We fit a blackbody model to the UV+U SED as we have done with previous nuclear transients (e.g., \citealt{hinkle20a}). We used Markov Chain Monte Carlo methods to find the best-fit blackblody parameters for the SED, using flat priors of $1000~\mathrm{K}~\leq~\mathrm{T}~\leq~55000 \mathrm~{K}$ and $10^{11}~\mathrm{cm}~\leq~\mathrm{R}~\leq~10^{17}~\mathrm{cm}$ in order to not overly influence the fits. The evolution in luminosity, effective radius, and temperature from these blackbody fits are shown in Figure \ref{fig:bbody_phasefold}. The blackbody models were poor fits if we included the $B$ and $V$ data. This could be due to deviations from a blackbody SED as a consequence of assuming the inter-flare flux is just the host galaxy. Since the UV dominates the energetics, using UV+U only should still give a reasonable characterization of the flares. For all flares observed in the UV, the peak blackbody luminosity and temperature occurred several days before the corresponding ASAS-SN $g$-band peaks. The blackbody radius begins to decline around the times of the $g$-band peaks. The peak blackbody luminosities of the flares were all consistent except for the July 2021 flare which was fainter but had a similar peak blackbody radius to the others. This is consistent with the luminosity evolution shown in Figure \ref{fig:phasefold_phot}. 

\vspace{10mm}

\subsection{TESS Light Curve Analysis}

\begin{figure*}\centering
\textbf{(a)}\includegraphics[width=\linewidth]{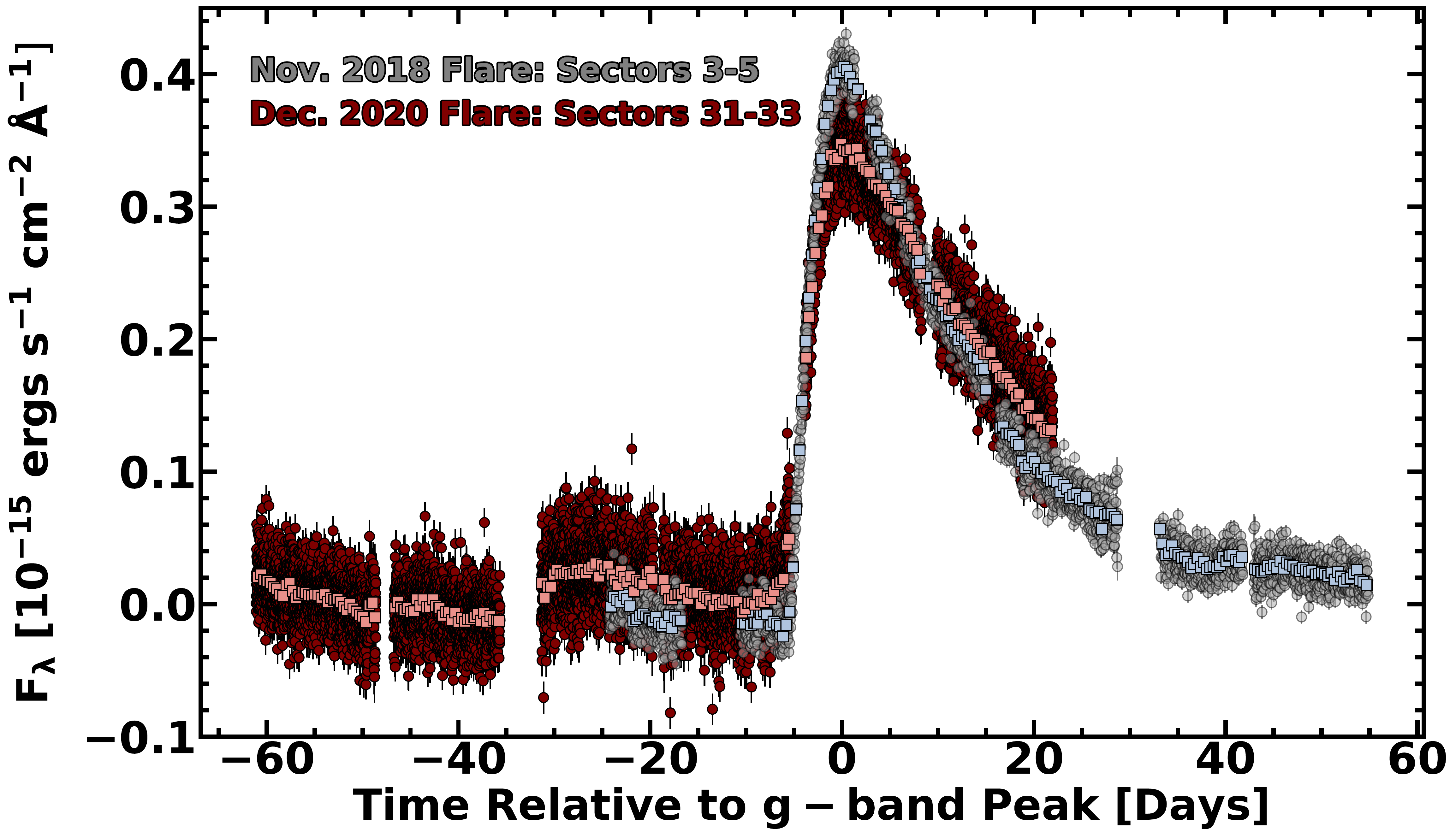}\hfill
\textbf{(b)}\includegraphics[width=.47\textwidth]{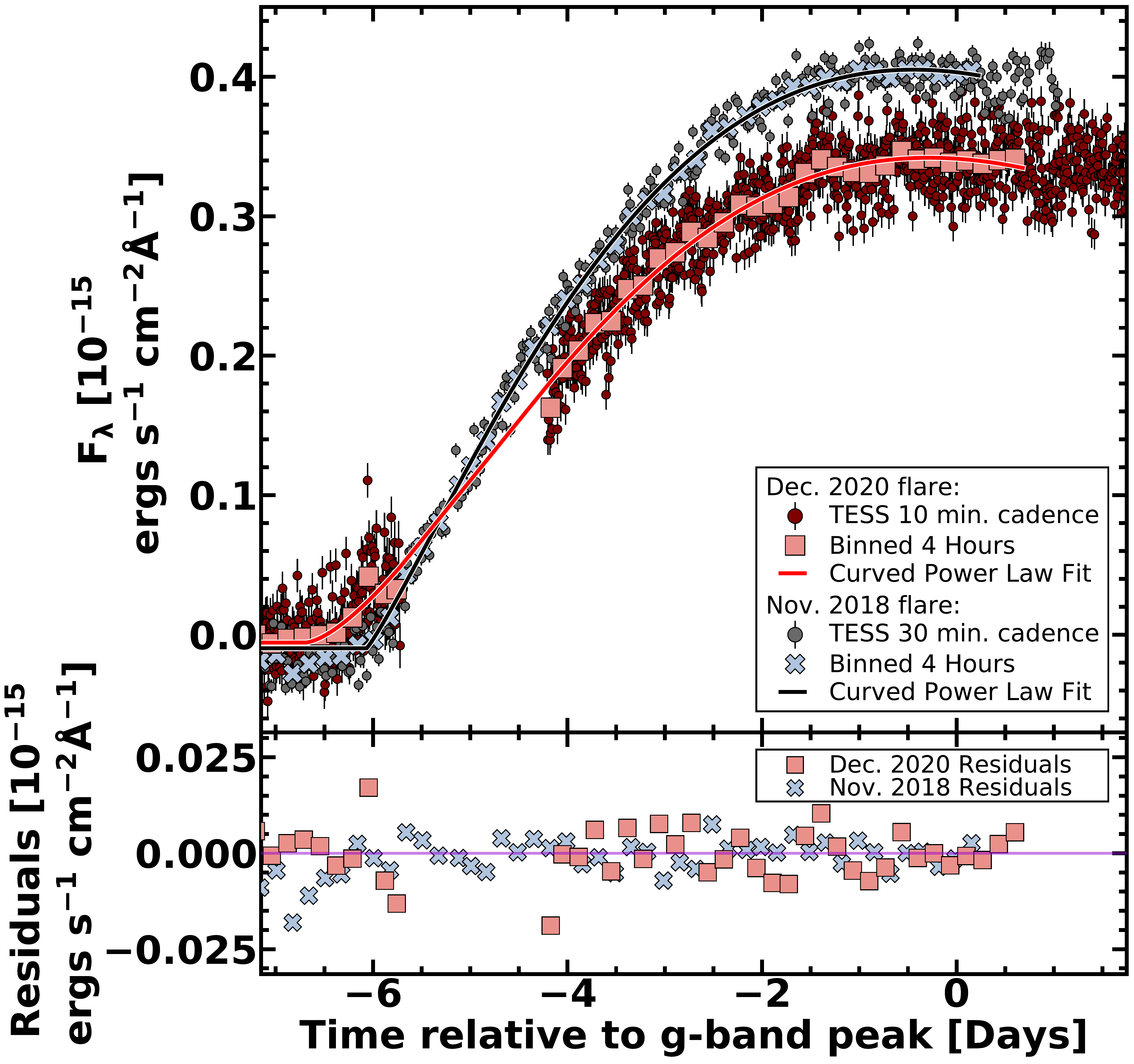}
\textbf{(c)}\includegraphics[width=.47\textwidth]{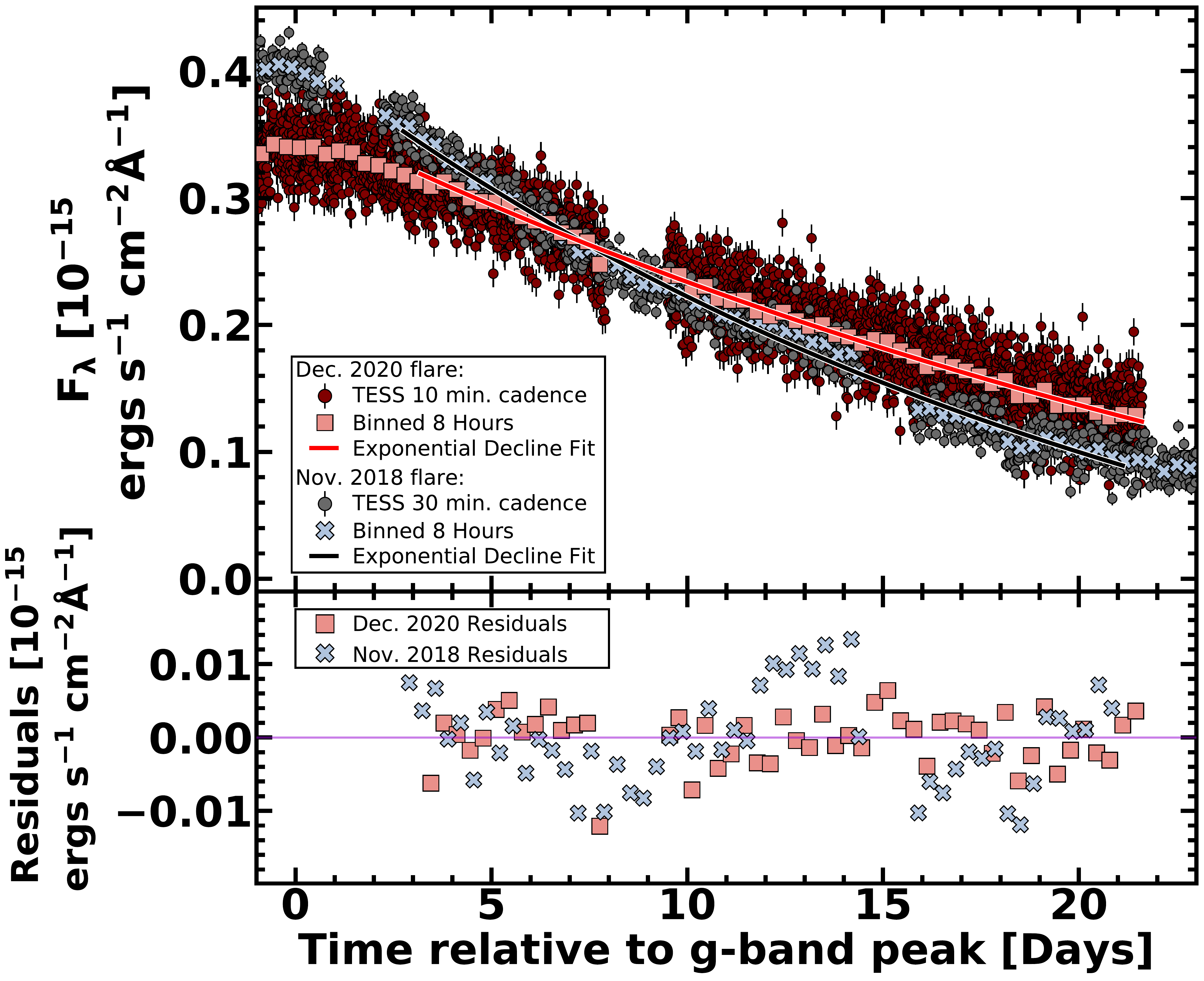}
\caption{(a) TESS image subtraction light curves of the December 2020 (red) and November 2018 (gray) flares. The two light curves are aligned using the updated timing model described in Section \ref{light_curve}. Squares show the data binned at 8 hour intervals. The unbinned December 2020 flare data has more scatter because of the increased cadence of the observations (10 minutes instead of 30 minutes). (b) The rising phase of the two flares with the corresponding best-fit curved power-law models (Equation \ref{eq:1}). The bottom panel shows the flux residuals. (c) The declining phase of the two flares with the best-fit exponential decline models shown in black and red for November 2018 and December 2020, respectively. For both (b) and (c) the \textit{TESS} data binned in 4 hour intervals are shown by squares and Xs in the color corresponding to the model fits.  }
\label{fig:total_tess_figure}
\end{figure*}

\textit{TESS} observed ASASSN-14ko's flares during Sectors 3-5, which captured the November 2018 flare (flare 12) \citep{payne2021a}, and again during Sectors 31-33 for the December 2020 flare (flare 18). \textit{TESS}'s second view of ASASSN-14ko's flaring behavior enables us to compare the properties of two individual flares using high-cadence data to investigate how they changed over that two year time frame. We aligned the data  using the updated timing model and plot this in Figure \ref{fig:total_tess_figure}. Due to the timing of the \textit{TESS} sectors, the full decline was not captured for the December 2020 flare, but the full rise to peak was observed. 

Previous TDEs and some supernovae have been fit by the model
\begin{equation}
    f = f_0 + f_1 (t-t_0)^a
\end{equation}
finding $\alpha \simeq 1.9-2.8$ (ASASSN-19bt with $\alpha = 2.10 \pm 0.12$, \citealt{holoien19}; ASASSN-19dj with $\alpha = 1.9 \pm 0.4$, \citealt{hinkle2021_29dj};  ZTF19abzrhg with $\alpha = 1.99 \pm 0.01$, \citealt{nicholl2020}).  In \citet{payne2021a}, we fit the first 25\% of the November 2018 rise with a single power-law to find $\alpha = 1.01 \pm 0.07$. There is a certain arbitrariness to these models in selecting the time period over which to do the fit because the actual light curve has to deviate from the power law as it approaches the peak.  The sector shift during the rise in December 2020 also makes it difficult to use this model.

Here we instead use the model used by \citet{vallely2021} to model the rise time of core-collapse supernovae with TESS data, 
\begin{equation} \label{eq:1}
    f(t) = \frac{h}{(1 + z)^2} \left( \frac{t - t_1}{1 + z} \right) ^{ a_1 \left( 1 + a_2 \dot (t-t_1) / (1+z) \right) } + f_0
\end{equation}
up to near the light curve peak for $t > t_1$ and as $f(t) = f_0$ for $ t < t_1$. Here $f_0$ is any residual background flux, $t_1$ is the beginning of the model rise, and the factors of $1+z$ are introduced to account for redshift time-dilation, although the low redshift of the host galaxy makes the correction small. Parameter $a_2$ is mathematically related to the rise time, $t_{rise} = t_{peak} - t_1$, between the start of the rise at time $t_1$ and the time of the peak $t_{peak}$. This model is able to adjust to the changing curvature as the transient approaches its peak and so can be fit over a broad enough time range for the slope of the initial rise to be unaffected by this choice (although the parameters for the curvature approaching peak will be). Using this model we find $a_1 = 1.10 \pm 0.04$ and $a_1 = 1.50 \pm 0.10$ for November 2018 and December 2020, respectively. The fits for both flare events using Equation \ref{eq:1} are shown in Figure \ref{fig:total_tess_figure}. The fit parameters are summarized in Table \ref{tab:tess_rise_decline_fit_params}. For our estimates of $t_{rise}$, we estimate $t_{peak}$ directly from the light curve using a polynomial fit to the data around peak. As shown by these model fits in Figure \ref{fig:total_tess_figure}, the December 2020 flare started to rise earlier and took a day longer to rise to peak than the November 2018 flare. In addition, the December 2020 flare took longer to rise to 50\% of the peak flux. We refit the previously-studied TDEs with publicly available photometric data with observed early-time rises with Equation \ref{eq:1} in order to compare to ASASSN-14ko and found $\alpha_1 = 5.12 \pm 0.12$ and $\alpha_2 = -0.02 \pm 0.001$ for ASASSN-19bt, and $\alpha_1 = 1.70 \pm 0.44$ and $\alpha_2 = -0.01 \pm 0.001$ for ASASSN-19dj. The early-time rise of ASASSN-14ko is similar to ASASSN-19dj but different than ASASSN-19bt.

Like the November 2018 flare \citep{payne2021a}, the December 2020 flare also smoothly declined in brightness, as shown in Figure \ref{fig:total_tess_figure}. As in \citet{payne2021a}, we fit the decline of both flares using a power-law and exponential model. Since only part of the decline was captured for the December 2020 flare, we fit both flares over this restricted time range (3 to 21 days after peak). The power-law model is 
\begin{equation}
    f(t) = z - h\left( \frac{t - t_{\mathrm{0}}}{\mathrm{days}} \right)^{\alpha}
\end{equation}
with $t_{\mathrm{0}}$ being the time of disruption, constrained to be before the start of the rise, $t_1$, determined above. However, in \citet{payne2021a} we found that an exponential decline   
\begin{equation}
    f(t) = a e^{- (t - t_{\mathrm{peak}} ) / \tau}  + c
\end{equation}
was a better fit for the November 2018 flare. We find that the December 2020 flare is also best fit by the exponential model with a reduced $\chi^2$ of 0.80 with 2419 degrees of freedom versus a reduced $\chi^2$ of 1.02 with 5851 degrees of freedom for the power-law model fit after inflating the errors to give the power-law model a reduced $\chi^2 \simeq 1$. There is a difference in the degrees of freedom because the power-law model also includes the pre-rise quiescence data as part of the fit whereas the exponential model does not. The best-fit parameters for both models are also shown in Table \ref{tab:tess_rise_decline_fit_params}. 

Compared to the November 2018 flare, the December 2020 flare began to rise earlier but took longer to rise to a fainter peak, and then declined more slowly. Although the full decline was not observed by \textit{TESS}, it is likely that the December 2020 flare also took longer to return to quiescence than the November 2018 flare.

\begin{table*}[htp!]
\centering
\caption{ Best-fit parameters for the peak, rise, and decline models of the \textit{TESS} light curves for the November 2018 flare and December 2020 flare.  }
\begin{tabular*}{0.78\textwidth}{l l l}
\toprule
& \textbf{November 2018 flare} & \textbf{December 2020 flare} \\ 
\toprule 
\textit{Peak Fit} & & \\
\hspace{5mm} $t_{peak}$ (JD) & 2458435.50$^{+0.13}_{-0.10}$ & 2459206.13$ \pm 0.10$ \\ 
\hspace{5mm} $F_{\lambda peak}$ (10$^{-15}$ ergs s$^{-1}$ cm$^{-2}$ $\AA^{-1}$) & $\hspace{9.5mm}0.40\pm 0.001$ & $\hspace{9.5mm}0.34 \pm 0.001$ \\
\hline
\textit{Curved Power Law Rise Fit} & & \\
\hspace{5mm} $t_1$ (JD) & $2458429.71 \pm 0.04 $  & $2459199.25 \pm 0.09$ \\
\hspace{5mm} $h$ (ergs s$^{-1}$ cm$^{-2}$ $\AA^{-1}$) & $\hspace{9.5mm}0.14 \pm 0.01$ & $\hspace{9.5mm}0.06 \pm 0.01$ \\
\hspace{5mm} $a_1$ & $\hspace{9.5mm}1.10 \pm 0.04$ & $\hspace{9.5mm}1.50 \pm 0.10$ \\
\hspace{5mm} $a_2$ (Days$^{-1}$) & $\hspace{7mm}-0.07 \pm 0.04$ & $\hspace{7mm}-0.06 \pm 0.10$ \\
\hspace{5mm} $f_0$ (ergs s$^{-1}$ cm$^{-2}$ $\AA^{-1}$) & $\hspace{7mm}-0.01 \pm 0.001$ & $\hspace{7mm}-0.01 \pm 0.001$ \\
\hspace{5mm} $t_{50}$ (days) & $\hspace{9.5mm}1.61 \pm 0.12 $ & $\hspace{9.5mm}2.49 \pm 0.13$ \\
\hspace{5mm} $t_{rise}$ (days) & $\hspace{9.5mm}5.80 \pm 0.12$ & $\hspace{9.5mm}6.88 \pm 0.13$ \\
\hline
\textit{Power Law Decline Fit} & & \\
\hspace{5mm} $t_0$ (JD) & $2458429.8 \pm 0.8 $ & $2459199.1 \pm 1.0$ \\
\hspace{5mm} $z$ (ergs s$^{-1}$ cm$^{-2}$ $\AA^{-1}$) & $\hspace{7mm}-0.01 \pm 0.001$ & $\hspace{7mm}-0.01 \pm 0.001$ \\
\hspace{5mm} $h$ (ergs s$^{-1}$ cm$^{-2}$ $\AA^{-1}$) & $\hspace{7mm}-3.5 \pm 0.7$ & $\hspace{7mm}-2.3 \pm 0.5$ \\
\hspace{5mm} $\alpha$ & $\hspace{7mm}-1.01 \pm 0.06$ & $\hspace{7mm}-0.81 \pm 0.05$ \\
\hline 
\textit{Exponential Decline Fit} & & \\
\hspace{5mm} $a$ (ergs s$^{-1}$ cm$^{-2}$ $\AA^{-1}$) & $\hspace{9.5mm}0.47 \pm 0.02$ & $\hspace{9.5mm}0.45 \pm 0.03$ \\
\hspace{5mm} $\tau$ (days) & $\hspace{8mm}22.3 \pm 1.2$ & $\hspace{8mm}33 \pm 3$ \\
\hspace{5mm} $c$ (days) & $\hspace{7mm}-0.12 \pm 0.02$ & $\hspace{7mm}-0.13 \pm 0.03$ \\

\toprule 

\end{tabular*}
\label{tab:tess_rise_decline_fit_params}
\end{table*}

\section{UV Spectroscopic Analysis} \label{uvspectra}

\begin{figure*}
    \centering
    \includegraphics[width=1\linewidth]{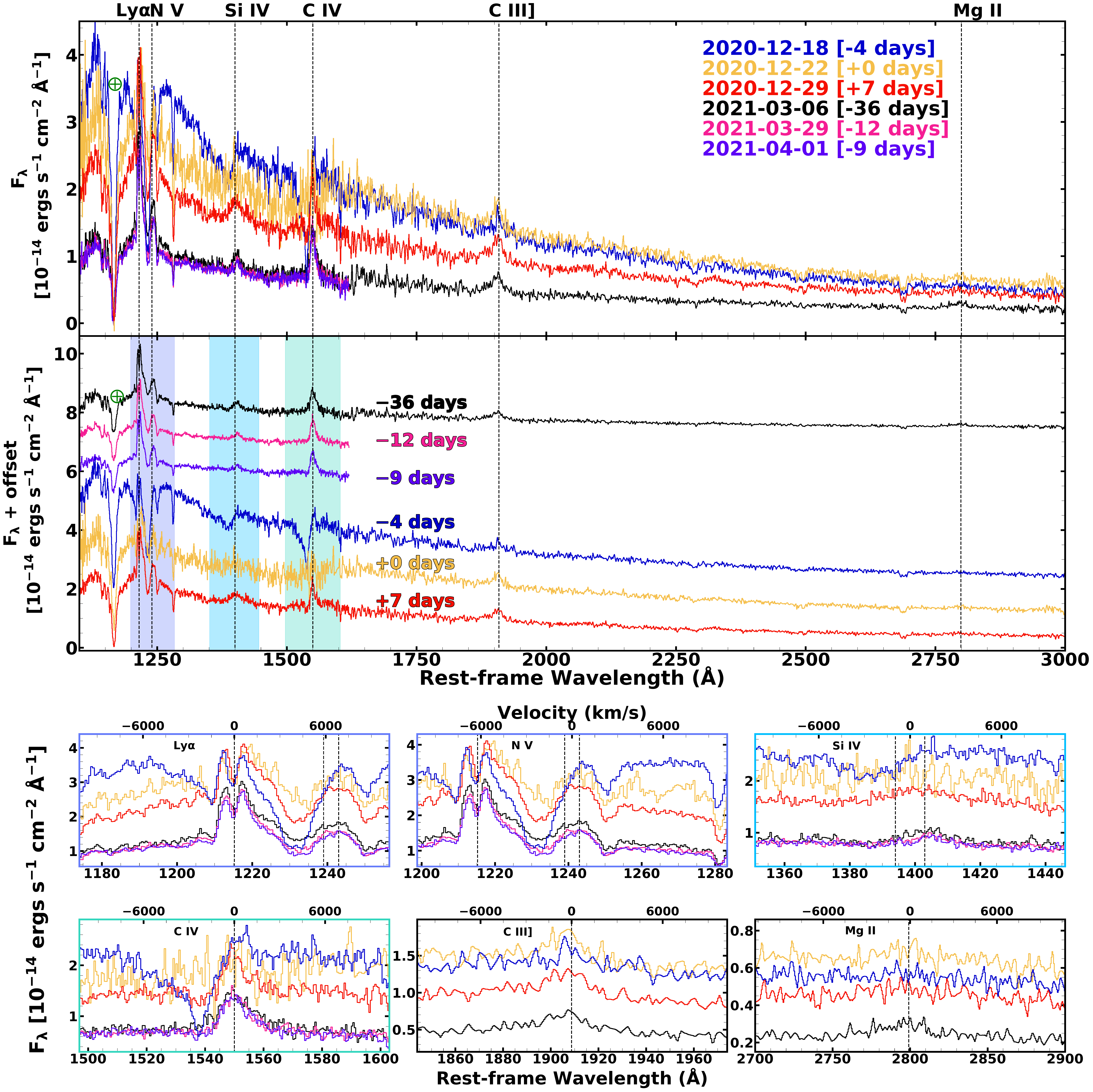}
    \caption{\textit{HST}/STIS UV spectra of ASASSN-14ko during the December 2020 and April 2021 flares. The $-4$ (blue), $+0$ (orange), and $+7$ (red) day spectra are relative to the December 2020 optical peak timing of MJD 59205.65, and the $-36$ (black), $-12$ (pink), and $-9$ day (purple) spectra are relative to the April 2021 optical peak timing of MJD 59315.22. The spectra have been corrected for Galactic extinction and the major geocoronal airglow lines are marked with the green $\oplus$ symbol. The middle panel shows the full spectra offset to better illustrate the change in line profiles over time. The bottom panels are zoomed-in to the labeled lines and not offset, showing the change in flux at different epochs of the flares. The major BALs are shaded in purple, blue, and green regions.   }
    \label{fig:all_STIS}
\end{figure*}

We present the six epochs of \textit{HST}/STIS UV Galactic extinction corrected spectra in Figure \ref{fig:all_STIS}. They were obtained during both the UV rise and decline for the December 2020 flare, as denoted by the color-coded vertical bars in Figure \ref{fig:phasefold_phot}. We also obtained two UV spectra a few days prior to the start of the UV rise of the April 2021 flare when the X-ray flux was anticipated to be in a low state based on the \textit{Swift} XRT light curves of the prior flares (see Section \ref{xray}). 

There were dramatic changes in both the UV continuum and the spectral lines as the December 2020 flare evolved. The FUV continuum was at its bluest state in the first observation $-4$ days prior to the optical peak and when the UV flux was still increasing. Several days later at epochs of +0 and +7 days, when the UV flux was declining, the UV continuum flattened. The continuum was faintest at $-36$, $-12$, and $-9$ days.  

The spectral lines during the flare are very different from the quiescent spectrum, as shown in Figure \ref{fig:all_STIS}, where we also offset each epoch to show the change in spectral features over time.  The prominent features are the standard UV lines of AGNs, including Ly$\alpha$, N V $\lambda 1240$, Si IV $\lambda \lambda 1394,1403$, C IV $\lambda 1550$, C III] $\lambda 1909$ and Mg II $\lambda2800$. In addition to these features, we detect narrow, high- and low-ionization absorption features at $z=0$ that we interpret as originating from the Milky Way ISM, as discussed in the Appendix. 

\begin{figure*}
    \centering
    \includegraphics[width=0.99\linewidth]{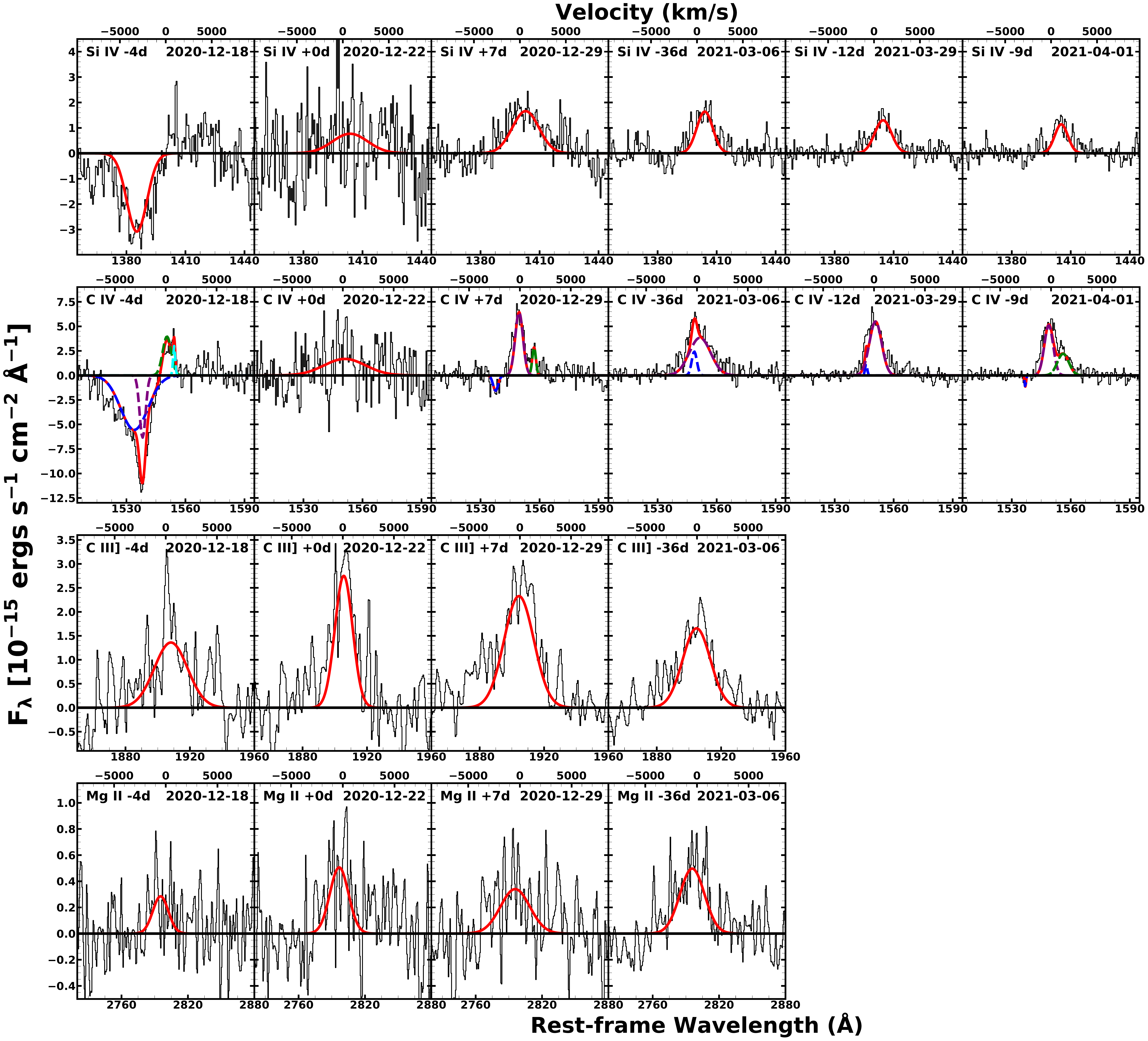}
    \caption{Continuum subtracted \ion{Si}{4}, \ion{C}{4}, \ion{C}{3}], and \ion{Mg}{2} profiles from the \textit{HST}/STIS UV spectra. The black lines show the spectra, the red lines show the best fitting Gaussian profiles, and the blue, purple, green and cyan dotted lines indicate the best fitting components of those profiles for \ion{C}{4}. }
    \label{fig:HST_model_fits}
\end{figure*}

The evolution of the spectral lines over the duration of the flare is most dramatic for Ly$\alpha$, N V $\lambda 1240$, Si IV $\lambda \lambda 1394,1403$, and C IV $\lambda 1550$. We fit either single or multiple-component Gaussian models to \ion{Si}{4}, \ion{C}{4}, \ion{C}{3}], and \ion{Mg}{2}, as shown in Figure \ref{fig:HST_model_fits} with the resulting fit parameters summarized in Table \ref{tab:uv_spectra_fits}. We estimated the local continuum by fitting a polynomial to the adjacent featureless regions surrounding the lines. Although it appears that Ly$\alpha$+\ion{N}{5} shows a similar trend in its profile evolution, we do not fit these profiles due to the complexity created by the blended lines and the telluric features. Velocities of each component were then obtained using these fits. Most components have velocities on the order of several thousand km/s.  During the UV rise at $-4$ days, a strong blue absorption feature is seen for the \ion{S}{4} and \ion{C}{4} lines which weakens and then vanishes during the declining phase of the UV light curve. The absorption feature is also absent in the quiescent spectrum. The presence of the absorption feature is clearest for the C IV $\lambda 1550$ and Si IV $\lambda \lambda 1394,1403$ lines because there are no other nearby lines. The velocity of this feature is $\sim4000-5000$ km/s. Based on the evolution of the C IV $\lambda 1550$ and Si IV $\lambda \lambda 1394,1403$ lines, the strong absorption feature centered at $1235~\mathrm{\AA}$ is possibly associated with the blue wing of N V $\lambda 1240$. As the absorption features vanish, broad emission lines of \ion{C}{4} and \ion{Si}{4} appear with similar velocity widths but $\sim$4,000 km/s redwards of the absorption features. \ion{C}{3}] and \ion{Mg}{2} are only seen in emission and show only modest changes. The presence of \ion{Mg}{2} is a significant difference from all previous UV spectra of TDEs. 

We obtained two UV spectra on 2021-03-29 and 2021-04-01, $-12$ and $-9$ days prior to the optical peak to determine the spectral properties at the time of the minimum X-ray emission. There were no large differences between these two spectra and the quiescent spectrum from 2021-03-06, aside from the appearance of a weak absorption feature on the blue side of \ion{C}{4} in the $-9$ day spectrum and at the X-ray minimum. This feature may be the beginning of the prominent blue absorption feature present at $-4$ days relative to peak.  There was a significant difference in the minimum X-ray luminosity between the two flares, with the December 2020 minimum being significantly fainter than the April 2021 minimum for both the total and soft X-ray luminosity, as discussed below.

\begin{table*}[htp!]
\centering
\caption{ ASASSN-14ko absorption and emission features and their evolution over time. $\lambda_0$ is the expected rest-frame wavelength of each ion whereas $\lambda_c$ is measured from the Gaussian fits of each component. Equivalent widths are negative for emission lines and positive for absorption lines.   }
\begin{tabular*}{0.99\textwidth}{l l l l l l }
\toprule
Ion & Phase \& Date Observed & $\lambda_0$ (rest $\mathrm{\AA}$)  & Central Velocity (rest km s$^{-1}$)  & Equivalent Width (rest $\mathrm{\AA}$) & FWHM (km s$^{-1}$) \\ 
\toprule 

\ion{Si}{4} & $-4$d, 2020-12-18 & $\lambda \lambda 1394,1403$  & $-2630 \pm 107$ & $\hspace{4.mm} 2.9^{+0.9}_{-0.6}$ & $3690 \pm 260$  \\
\ion{Si}{4} & +0d, 2020-12-22 & $\lambda \lambda 1394,1403$  & $\hspace{2.5mm}1310 \pm 1390$ & $\hspace{1.5mm} -3.7^{+1.9}_{-2.4}$ & $6370 \pm 3260$  \\
\ion{Si}{4} & +7d, 2020-12-29 & $\lambda \lambda 1394,1403$  & $\hspace{2.5mm}1130 \pm 197$ & $\hspace{1.5mm} -3.9^{+0.5}_{-0.5}$ & $4820 \pm 460$  \\
\ion{Si}{4} & $-36$d, 2021-03-06 & $\lambda \lambda 1394,1403$  & $\hspace{2.5mm}1360 \pm 116$ & $-13.1^{+4}_{-8}$ & $2900 \pm 270$  \\
\ion{Si}{4} & $-12$d, 2021-03-29 & $\lambda \lambda 1394,1403$  & $\hspace{2.5mm}1500 \pm 116$ & $\hspace{1.5mm} -3.3^{+0.2}_{-0.3}$ & $3040 \pm 270$  \\
 \ion{Si}{4} & $-9$d, 2021-04-01 & $\lambda \lambda 1394,1403$  & $\hspace{2.5mm}1620 \pm 122$ & $\hspace{1.5mm} -2.5^{+0.3}_{-0.3}$ & $2380 \pm 290$  \\
\hline
\ion{C}{4} & $-4$d, 2020-12-18 & $\lambda1550$  & $-2990 \pm 367$ & $\hspace{4mm} 8.3_{-1.4}^{+1.3}$ & $4500 \pm 510$  \\
  & & & $-2180 \pm 40.5$ & $\hspace{4mm} 2.2_{-0.5}^{+0.8}$  & $\hspace{1.5mm}980 \pm 140$ \\
 & & & $\hspace{4mm}212 \pm 154$ & $\hspace{1.5mm} -1.7_{-1}^{+0.8}$ & $1410 \pm 500$ \\
 & & & $\hspace{4mm}922 \pm 61.7$ & $\hspace{1.5mm} -0.4_{-0.5}^{+0.2}$ & $\hspace{1.5mm}340 \pm 190$ \\
\ion{C}{4} & +0d, 2020-12-22 & $\lambda1550$  & $\hspace{4mm}289 \pm 965$ & $\hspace{1.5mm} -7.7_{-2.1}^{+2.0}$ & $6810 \pm 2270$ \\
\ion{C}{4} & +7d, 2020-12-29 & $\lambda1550$  & $-2280 \pm 118$ & $\hspace{4mm} 0.7^{+0.2}_{-0.2}$ & $\hspace{1.5mm}840 \pm 280$ \\
 &  &   & $\hspace{3.5mm}15.4 \pm 46.3$ & $\hspace{1.5mm} -4.0_{-0.4}^{+0.5}$ & $1270 \pm 110$ \\
 &  &   & $\hspace{2.5mm}1490 \pm 69.5$ & $\hspace{1.5mm} -0.9_{-0.8}^{+0.3}$ & $\hspace{1.5mm}590 \pm 160$ \\
\ion{C}{4} & $-36$d, 2021-03-06 & $\lambda1550$  & $\hspace{1.5mm}-164 \pm 84.9$ & $\hspace{1.5mm} -2.0_{-1.2}^{+0.7}$  & $\hspace{1.5mm}850 \pm 240$ \\ 
 &  &   & $\hspace{4mm}390 \pm 106$ & $-13.2_{-1.1}^{+1.5}$ & $3470 \pm 250$ \\ 
\ion{C}{4} & $-12$d, 2021-03-29 & $\lambda1550$  & $\hspace{1.5mm}-675 \pm 77.2$ & $\hspace{1.5mm} -0.5_{-0.3}^{+0.4}$  & $\hspace{1.5mm}330 \pm 210$ \\ 
 &  &   & $\hspace{4mm}243 \pm 50.2$ & $-11.6_{-0.7}^{+0.9}$  & $2000 \pm 100$ \\ 
\ion{C}{4} & $-9$d, 2021-04-01 & $\lambda1550$  & $-2440 \pm 48.2$ & $\hspace{4mm}  0.3_{-0.10}^{+0.13}$  &  $\hspace{1.5mm}240 \pm 120$ \\ 
 &  &   & $\hspace{1.5mm}-135 \pm 174$ & $\hspace{1.5mm} -8.3_{-1.6}^{+2.1}$  & $1440 \pm 210$ \\ 
 &  &   & $\hspace{2.5mm}1270 \pm 598$ & $\hspace{1.5mm} -4.3_{-2}^{+1.7}$  & $1860 \pm 770$ \\ 
\hline
\ion{C}{3}] & $-4$d, 2020-12-18 & $\lambda1909$  & $\hspace{0.5mm}-78.3 \pm 219$ & $\hspace{1.5mm} -5.9_{-0.5}^{+0.4}$ & $5390 \pm 520$ \\
\ion{C}{3}] & +0d, 2020-12-22 & $\lambda1909$  & $\hspace{1.5mm}-490 \pm 86.2$ & $\hspace{1.5mm} -7.2_{-0.4}^{+0.5}$ & $2870 \pm 200$ \\
\ion{C}{3}] & +7d, 2020-12-29 & $\lambda1909$  & $\hspace{1.5mm}-670 \pm 95.6$ & $-13.9_{-0.7}^{+0.6}$ & $4840 \pm 220$ \\
\ion{C}{3}] & $-36$d, 2021-03-06 & $\lambda1909$  & $\hspace{1.5mm}-620 \pm 97.1$ & $-19.8_{-1.1}^{+1.0}$ & $4510 \pm 230$ \\
\hline
\ion{Mg}{2} & $-4$d, 2020-12-18 & $\lambda2800$  & $\hspace{1.5mm}-406 \pm 278$ & $\hspace{1.5mm} -1.4^{+0.3}_{-0.3}$ & $2490 \pm 660$ \\
\ion{Mg}{2} & +0d, 2020-12-22 & $\lambda2800$  & $\hspace{1.5mm}-256 \pm 192$ & $\hspace{1.5mm} -2.1^{+0.3}_{-0.3}$ & $3000 \pm 450$ \\
\ion{Mg}{2} & +7d, 2020-12-29 & $\lambda2800$  & $\hspace{1.5mm}-363 \pm 267$ & $\hspace{1.5mm} -2.8_{-0.4}^{+0.3}$ & $4740 \pm 630$ \\
\ion{Mg}{2} & -36d, 2021-03-06 & $\lambda2800$  & $\hspace{1.5mm}-395 \pm 128$ & $\hspace{1.5mm} -7.0_{-0.7}^{+0.6}$ & $3950 \pm 300$ \\

\toprule 

\end{tabular*}
\label{tab:uv_spectra_fits}
\end{table*}

\vspace{10mm}

\section{Analysis of the X-ray Emission} \label{xray}
As previously shown in \citet{payne2021a,payne2021b}, ASASSN-14ko's X-ray evolution during the flares differs from its UV/optical evolution. The X-ray luminosity rapidly drops during the UV/optical rise, recovers near peak and then declines again before recovering. Here we discuss the new X-ray data from \textit{Swift} and {NICER} during December 2020, April 2021, July 2021, and November 2021. 

\subsection{X-ray Light Curve from Swift XRT and NICER}
The X-ray light curves and their hardness ratios are compared to the UV/optical light curves in Figure \ref{fig:full_LC} and the aligned light curves are shown in Figure \ref{fig:phasefold_phot}. The X-ray luminosity follows the previously observed trends. The X-ray luminosity consistently has a minimum ${\sim}5$ days before the optical peak but the depth of the minimum is variable. The largest decreases were for the September 2020 and December 2020 flares, and the smallest decrease was for the April 2021 flare. 

\begin{figure*}
    \centering
    \includegraphics[width=0.9\linewidth]{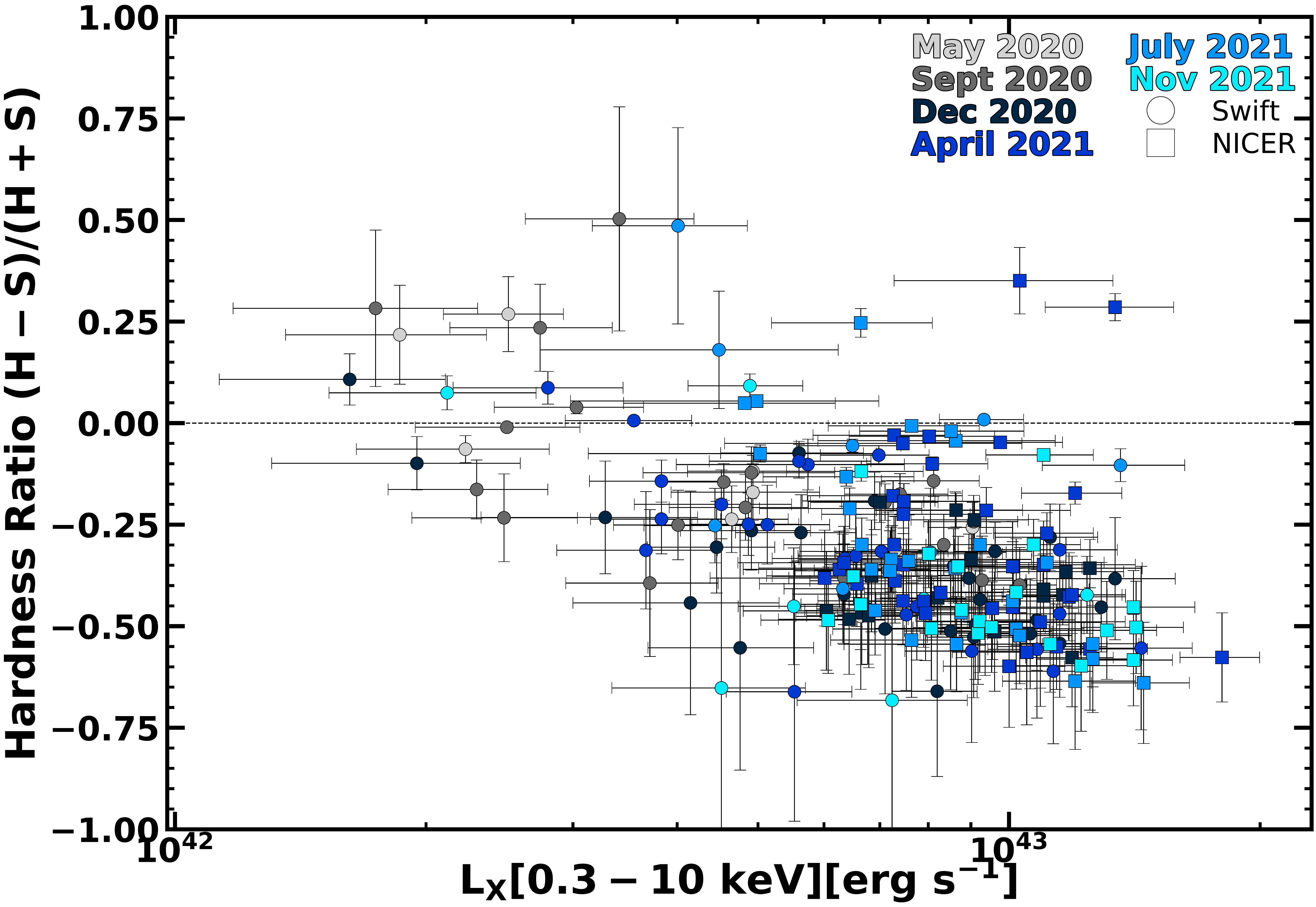}
    \caption{The connection between the hardness ratio evolution and the X-ray luminosity for all ASASSN-14ko flares observed with \textit{Swift} XRT and NICER.   }
    \label{fig:xray_lum_HR_evol}
\end{figure*}

\begin{figure*}
    \centering
    \includegraphics[width=\linewidth]{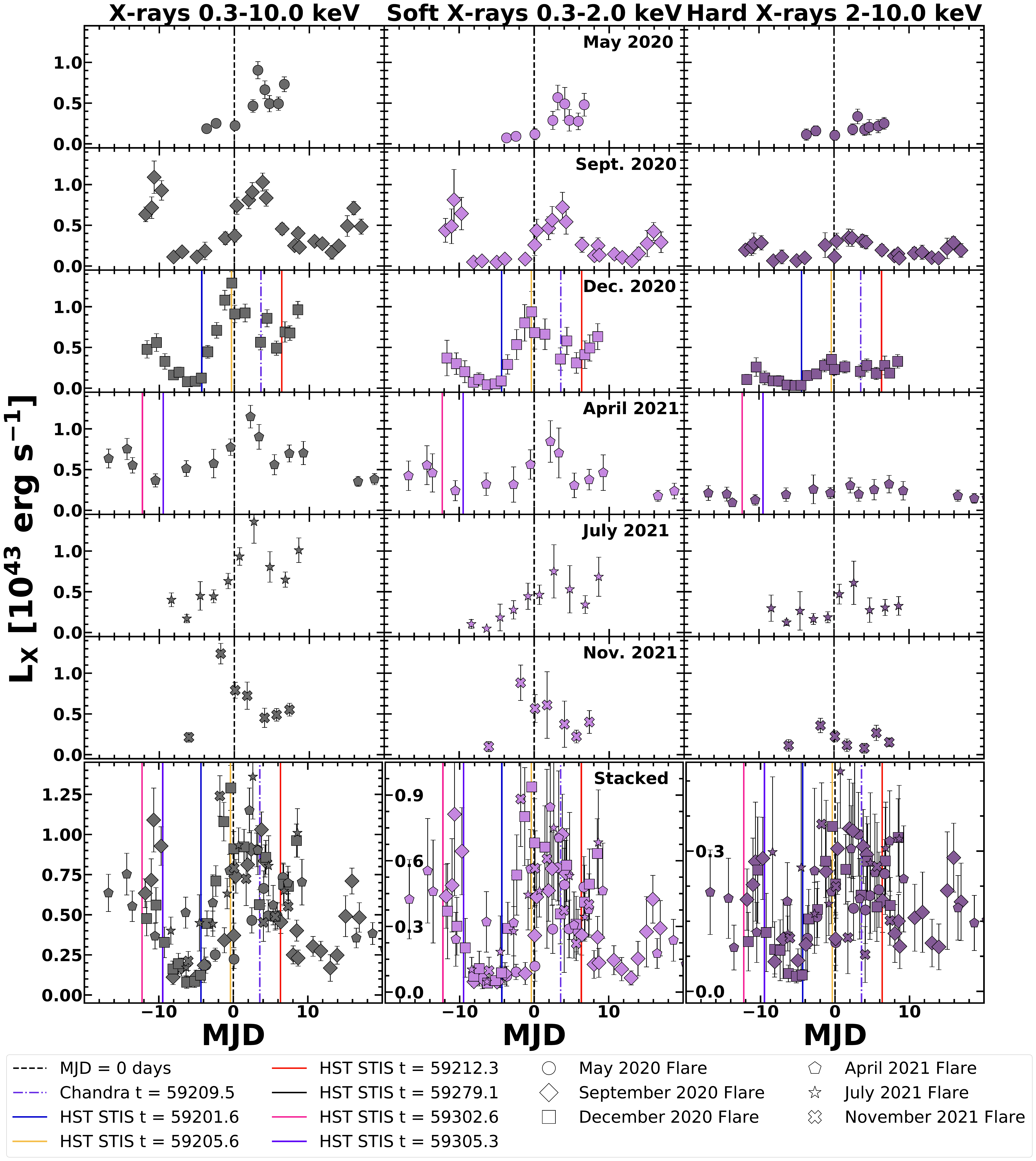}
    \caption{\textit{Swift} XRT X-ray luminosities separated into total luminosity between $0.3-10.0$ keV, hard X-rays between $2-10.0$ keV, and soft X-rays between $0.3-2.0$ keV. The bottom panels in each column show the data stacked using the timing model.  }
    \label{fig:xrays_hard_soft_comparison}
\end{figure*}

Near the minimum, the hardness ratio (HR) also changes. During quiescence before the flares, the HR is ${\sim}-0.5$ and it then hardens to HR${\sim}0.1 - 0.5$ at the minimum. There was a secondary X-ray minimum at $+10$ days only in the September 2020 flare. This secondary minimum also had a harder spectrum with HR $\sim0.25$. This ``softer when brighter, harder when fainter" behavior is illustrated in Figure \ref{fig:xray_lum_HR_evol}. Across all flares, both the hard and soft X-ray luminosities dip together, as shown in Figure \ref{fig:xrays_hard_soft_comparison}, but the soft X-rays consistently nearly disappear completely during the minimums. The April 2021 flare was unusual in that both the total X-ray and soft X-ray luminosity were unusually bright compared to the other flares. This indicates that the X-rays were uncharacteristically bright when the UV spectra were observed at $-12$ and $-9$ days relative to the April 2021 flare optical peak. Overall, Figure \ref{fig:xrays_hard_soft_comparison} shows that ASASSN-14ko's evolution is strongest in the softest energy band. 

\subsection{X-ray Spectra from Chandra}
\begin{figure*}
    \centering
    \textbf{(a)}\includegraphics[width=0.47\linewidth]{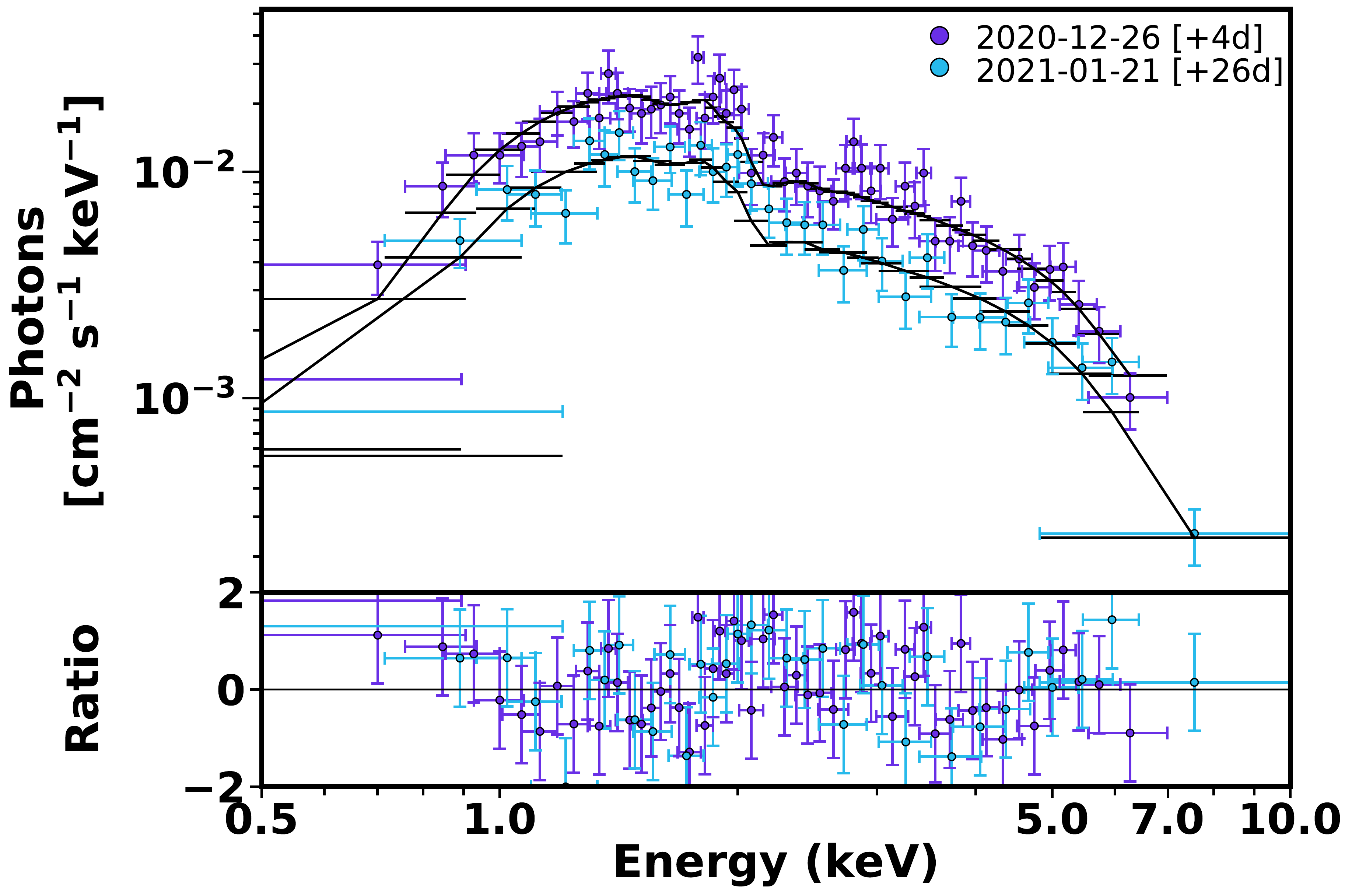}
    \textbf{(b)}\includegraphics[width=0.47\textwidth]{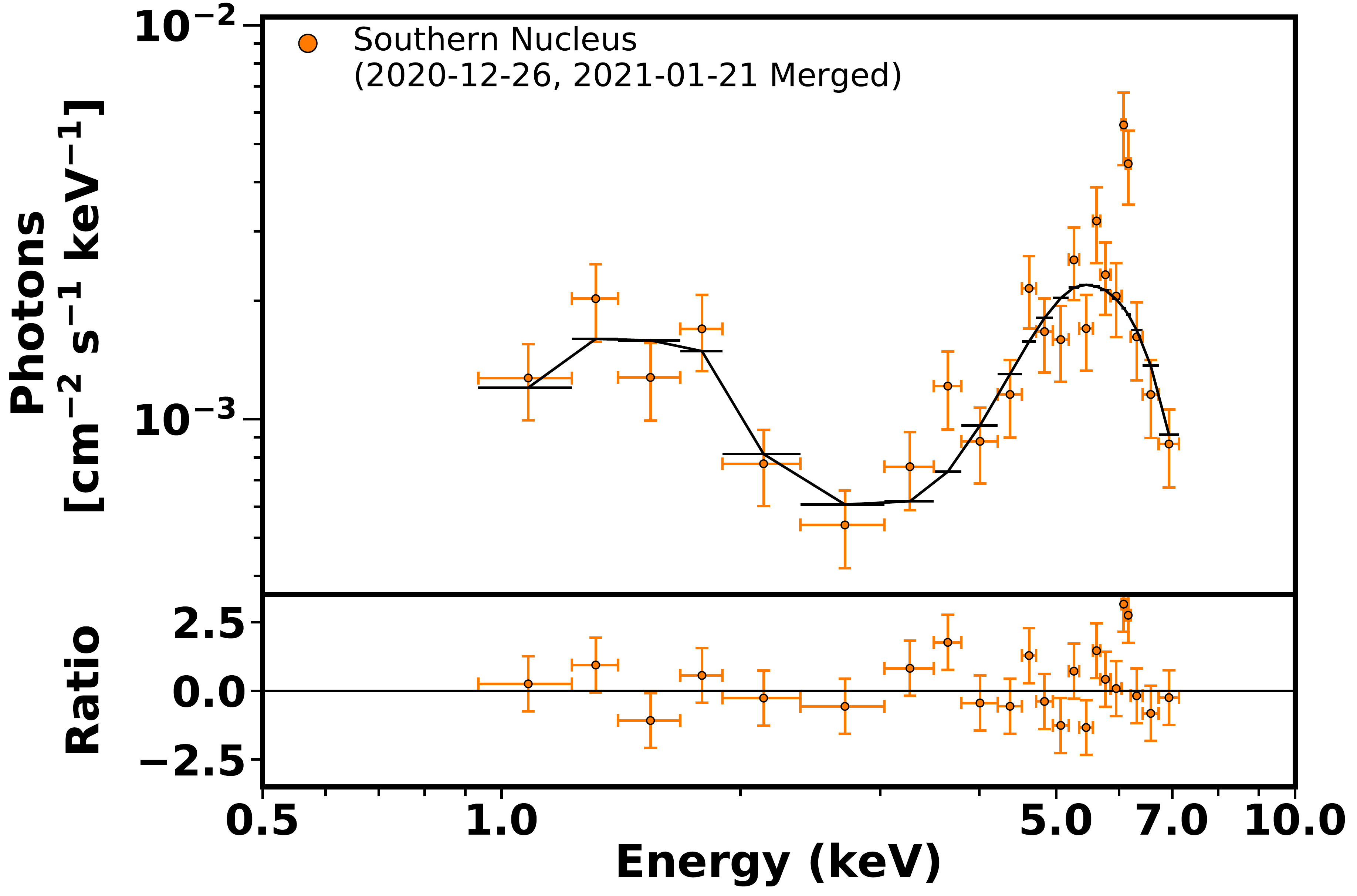}
    \textbf{(c)}\includegraphics[width=0.55\textwidth]{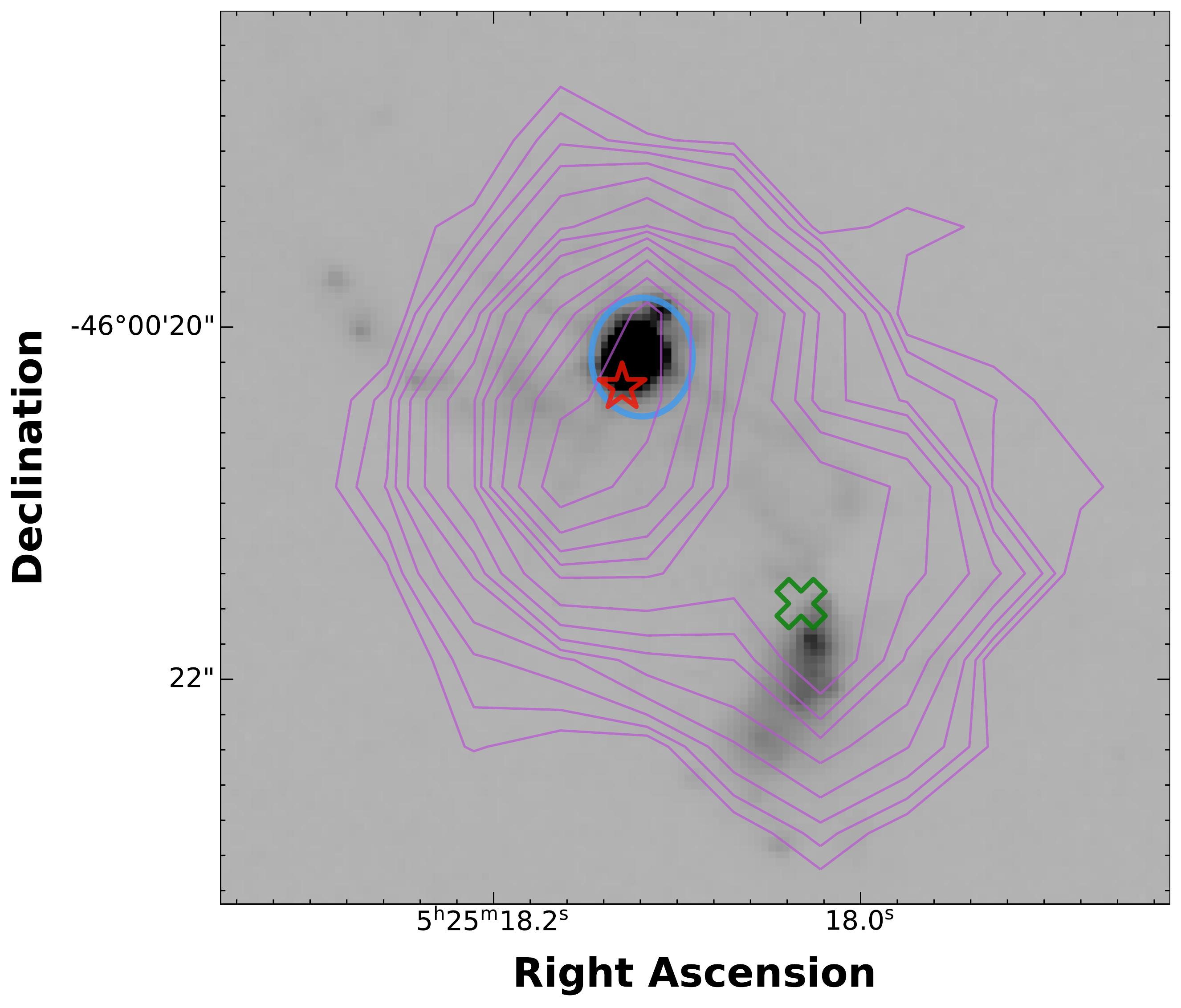}
    \caption{\textit{Chandra} X-ray spectra of \gal{}'s northern nucleus \textbf{(a)}, which corresponds to ASASSN-14ko, taken $+4$ days and $+31$ days after the $g$-band peak associated with the December 2020 flare, and \gal{}'s southern nucleus \textbf{(b)}. \textit{HST} WFC3 F275W UV image of \gal{} (PI: J. Lyman, Proposal ID: 16287) \textbf{(c)} overlaid with contours of the $0.3-10.0$ keV \textit{Chandra} image. The blue ellipse denotes the position corresponding to the location of the flares reported in \citet{payne2021b}, the red star shows the position originally reported for ASASSN-14ko by \citet{holoien14ATELc}, and the green X represents the position of the southwestern nucleus from \citet{tucker2020}. }
    \label{fig:chandra_fig}
\end{figure*}

Due to the spatial resolution of \textit{Swift} XRT and \textit{NICER}, it was not possible to separately analyze the X-ray properties of the two AGN nuclei present in \gal{}. As shown by the contours in Figure \ref{fig:chandra_fig}, \textit{Chandra} detects the two nuclei as separate sources coincident with the two nuclei identified by \citet{tucker2020} in the optical.  

We extracted X-ray spectra for both sources for the two epochs observed at $+4$ and $+26$ days relative to the December 2020 $g$-band peak. For both epochs, the absorbed power-law \texttt{xspec} model $\textrm{tbabs} \times \textrm{zashift} \times \textrm{powerlaw}$ with a photon index of $1.34\pm0.11$ at $+4$ days and $1.32\pm0.09$ at $+26$ days provided the best fit for ASASSN-14ko. These two spectra and the models are shown in Figure \ref{fig:chandra_fig}. Comparing these two spectra indicates that the northern nucleus was brighter at $+4$ days during the UV/optical light curve decline versus at $+26$ days. We also extracted spectra for the southern nucleus at each epoch but found that the fit was not well constrained. The merged spectrum combining both epochs was best fit as an absorbed power law with a photon index of $1.61^{+0.65}_{-0.58}$, as shown in Figure \ref{fig:chandra_fig}. The spectral properties are summarized in Table \ref{tab:chandra_spectra_fits}.  The Fe K$\alpha$ line seen in the archival \textit{XMM-Newton} and \textit{NuSTAR} observations \citep{payne2021a} originates from the southern nucleus and not from ASASSN-14ko. 

\begin{table*}[htp!]
\centering
\caption{ Best-fit parameters for absorbed power-law models of the \textit{Chandra} X-ray spectra of the northern and southern nucleus. The first observation had 55 degrees of freedom and the second observation had 31 degrees of freedom for the northern nucleus. The southern spectrum was produced by merging the two epochs, and this spectrum was fit with an additional Gaussian with energy $6.57^{+0.26}_{-0.21}$ keV and width $1.25_{ -0.21}^{+0.27}$ keV.    }
\begin{tabular*}{\textwidth}{l l l l l l l}
\toprule
 MJD & MJD-MJD$_{\mathrm{g ~ peak}}$ & \pbox{25mm}{N$_{\mathrm{H}}$ \\ (10$^{22}$ g$^{-1}$cm$^{-1}$)} &  \pbox{43mm}{Absorbed Flux (0.3-10.0 keV) \\ (erg/cm$^{2}$/s)} & \pbox{7mm}{ log$_{10}$(L$_{\mathrm{x}}$) \\ (erg/s) } & Photon Index $\Gamma$ & $\chi^2$ per dof \\ 
\hline
\textit{Northern:} \\
\hspace{5mm} 59210.0 & $+4.4$ & $0.002\pm{0.07}$ & (1.4$_{-0.2}^{+0.3}$) $\times 10^{-12}$ & 42.77$_{-0.07}^{+0.08}$ & $1.34\pm0.11$ & 0.68 \\

\hspace{5mm} 59236.3 & $+30.7$ & $0.002\pm0.10$ & (7.6$_{-1.8}^{+1.7}$) $\times 10^{-13}$ & 42.50$_{-0.11}^{+0.10}$ & $1.32\pm0.09$ & 0.84 \\

\textit{Southern:} \\
\hspace{5mm} ... & ... & 0.035 & (5.8$_{-0.5}^{+0.5}$) $\times 10^{-13}$ & 42.38$_{-0.04}^{+0.04}$ & $1.61^{+0.65}_{-0.58}$ & 1.75 \\

\hline 
\end{tabular*}
\label{tab:chandra_spectra_fits}
\end{table*}

\begin{figure*}
    \centering
    \includegraphics[width=0.83\linewidth]{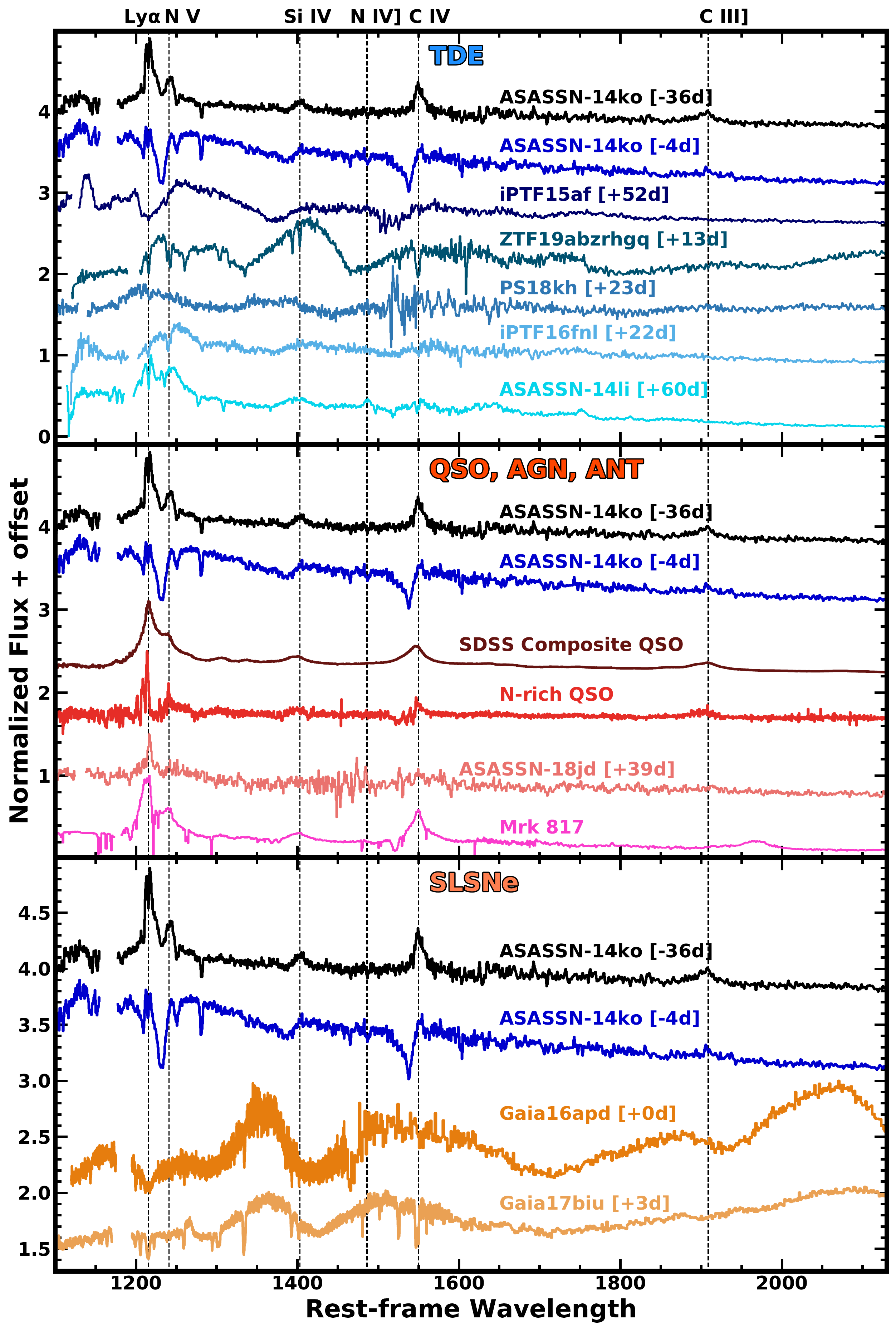}
    \caption{Comparison of the \textit{HST}/STIS UV spectrum of ASASSN-14ko taken four days prior to the December 2020 optical peak to other UV spectra. The top panel shows the TDEs iPTF15af \citep{blagorodnova2019}, ZTF19abzrhgq \citep{hung2020}, PS18kh \citep{hung2019}, iPTF16fnl \cite{brown2018}, ASASSN-14li \citep{cenko2016}. The middle panel shows the composite QSO spectrum of \citet{vandenberk2001}, the nitrogen rich QSO SDSS J164148.19+223225.2 \citep{batra2014}, the ambiguous nuclear transient (ANT) ASASSN-18jd \citep{neustadt2020}, and the AGN Mrk 817 \citep{kara2021}. The bottom panel shows the superluminous supernovae Gaia16apd \citep{yan2017}, and Gaia17biu \citep{yan2018}. All spectra are normalized and offset for visibility. The ordering of the spectra relative to ASASSN-14ko is qualitative. The geocoronal airglow emission in each spectra is masked.  }
    \label{fig:UV_comp}
\end{figure*}

\section{Discussion} \label{discussion}

We observed four of the most recent flares from ASASSN-14ko from X-ray through optical wavelengths. At the time of ASASSN-14ko's discovery as a periodic transient, ASAS-SN had observed sixteen flares over six years in the optical. Now, six sequential flares have been observed at X-ray, UV, and optical wavelengths.  

The six flares are UV dominated and have the blackbody SEDs typical of TDEs but with possible deviations going into the optical. They have different peak luminosities, indicating that the source is evolving, which was not clear from the lower quality ground-based light curves. The high-cadence \textit{TESS} data further show that the optical flares do differ in terms of the peak luminosity and rise/decline morphology, but the changes were more subtle than in the UV. The July 2021 flare's much fainter UV flux and flatter evolution indicate that the UV properties can vary widely. It remains to be seen if this was a unique or particularly rare event. 

Detecting these differences is important for understanding the physical nature of the system. For example, \citet{payne2021a} argued against interpreting the flares as being due to a star passing through and disrupting the SMBH accretion disk because the even and odd flares  (i.e., up/down through the disk) were indistinguishable given the quality of the data. Thus far, the six flares observed in the UV also do not have any obvious differences between even and odd flares.

For the TDE hypothesis, the flares have to evolve, and the flares have to eventually ($\sim100$ years based on $P_0 = 115.2^{+1.3}_{-1.2}$ days and $\dot{P} = -0.0026 \pm 0.0006$) come to an end.  In this scenario, the envelope of a giant star is steadily being stripped at each pericentric passage.  This means that the star has to evolve and the amount of mass stripped should change with time both secularly and stochastically.  The orbit should change with each pericentric passage, but it should not be completely regular -- the period derivative should vary, and we may see evidence for this in our inability to fit all the peak times with a single $\dot{P}$ (see Figure \ref{fig:o-c_plot}).  Recent theoretical work on modeling ASASSN-14ko has also supported the partial TDE hypothesis. \citet{cufari2022} used analytic arguments and three-body integrations to show that the Hills mechanism can result in the capture of a star on an orbit similar to ASASSN-14ko's observed period. \citet{metzger2022} proposed ASASSN-14ko as a system of two stars co-orbiting on an extreme mass ratio inspiral (EMRI) resulting in a long-lived mass-transfer. \citet{king2022} propose that ASASSN-14ko is caused by a white dwarf of mass $0.58M_{\odot}$ on $114$ day orbit but the tidal disruption radius of a white dwarf is deep inside the event horizon given the black hole mass estimates.

\begin{figure*}
    \centering
    \includegraphics[width=\linewidth]{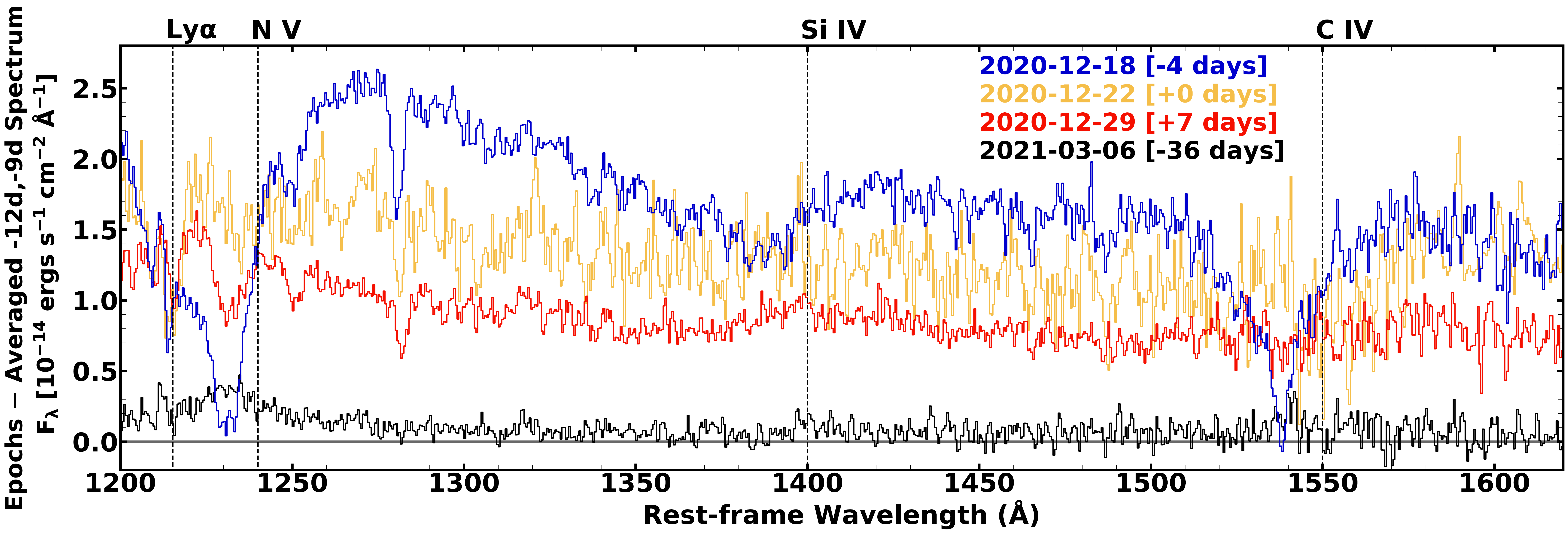}
    \caption{The \textit{HST}/STIS spectra epochs at $-4$, $+0$, $+7$, and $-36$ days after subtracting off an averaged spectra combining the $-12$ and $-9$ day epochs. The broad absorption features are optically thick in the core of the line. }
    \label{fig:UV_residual_spectra}
\end{figure*}

The strong UV spectral evolution is notable not only for the remarkable evolution of the broad absorption lines, but also for the extremely rapid timescale over which the line profiles changed. Compared to the UV spectra of both TDEs and AGNs, ASASSN-14ko's UV spectra changes in days versus the timescales of months or years observed for normal TDEs and changing-look AGNs. The general trend of the UV lines is the formation of a strong blue absorption feature at $\sim-5000$ km/s relative to the expected rest-frame central wavelength that appears prior to the optical peaks, and then vanishes and only a broad emission line around the rest-frame wavelength remains. This trend is most apparent for the \ion{C}{4} line, which shows the start of the formation of the blue absorption feature at $-9$ days and evolves into an emission line by $+7$ days. All the absorption features are deep but they are not saturated. For \ion{Si}{4} at $-4$ days, the line absorbs 17\% of the continuum flux at maximum depth and 68\% for \ion{C}{4} at $-4$ days. This implies that the absorbing material must both have a high optical depth and a high covering fraction. Speculatively, if the $-12$ day and $-9$ day spectra represent the AGN UV emission which is unchanging during the flare, then we can subtract these from the $-4$, 0 and $+7$ day flaring spectra as shown in Figure \ref{fig:UV_residual_spectra}. Then we find that the \ion{C}{4} and \ion{Si}{4} absorption is optically thick in the core of the line.  

\begin{figure*}
    \centering
    \includegraphics[width=\linewidth]{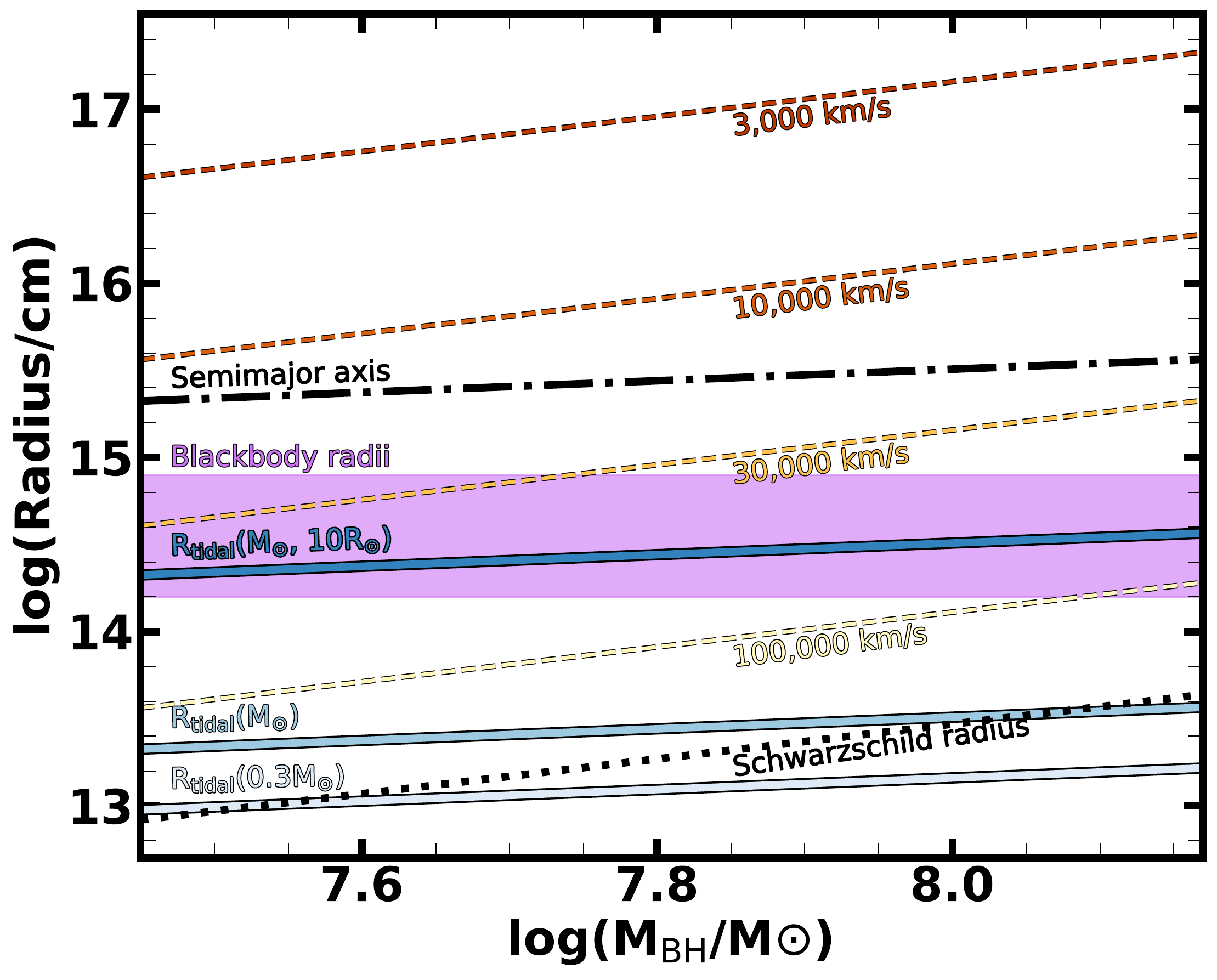}
    \caption{Radii as a function of SMBH mass $M_{BH}$ for the tidal radii corresponding to a $0.3M_{\odot}$ main-sequence star, the Sun, and a star with the mass of the Sun and a radius of $10R_{\odot}$ ($R_{\mathrm{tidal}}(0.3M_{\odot})$, $R_{\mathrm{tidal}}(M_{\odot})$, $R_{\mathrm{tidal}}(M_{\odot}, 10R_{\odot})$ shown as solid lines); Schwarzschild radius (dotted line); semimajor axis assuming $P = 115.2$ days (dash-dotted line);  radii corresponding to a range of circular velocities (dashed lines); and blackbody radii spanning $10^{14.2}$ cm to $10^{14.9}$ cm taken from Figure \ref{fig:bbody_phasefold} (shaded). The $\log_{10}(M_{BH})$ shown for both plots matches the range reported in \citet{payne2021a}.}
    \label{fig:BH_scales_R_V}
\end{figure*}

We compare the UV spectra during ASASSN-14ko's UV rise at $-4$ days and quiescence at $-36$ days to the UV spectra of TDEs, AGNs, and superluminous supernovae (SLSNe) in Figure \ref{fig:UV_comp}. The UV continuum of ASASSN-14ko at $-4$ days most closely resembles the TDEs rather than the AGNs, which have consistently flat continua from the FUV to the NUV. There are no similarities between ASASSN-14ko and SLSNe. Overall, ASASSN-14ko at $-4$ days most closely resembles iPTF15af, since both objects have broad \ion{N}{5}, \ion{Si}{4}, and \ion{C}{4} absorption features at similar velocities along with a blue continuum. ZTF19abzrhgq also showed broad absorption features around these lines but at much higher velocities of $\sim 15,000$ km/s \citep{hung2020}. However, \ion{C}{4} and \ion{Si}{4} in ASASSN-14ko appear similar to classic P Cygni profiles that are not apparent in any other TDE spectra. In comparison, the spectrum at $-36$ days is dominated by emission lines akin to AGNs. The strong broad absorption features observed at $-4$ days but not at $-9$ days could either indicate that the feature only forms several days before the optical peak or that they only form when the X-ray luminosities reach a very low state like in December 2020. 

The \ion{Mg}{2} emission line is typically present in AGNs and comes from a partially ionized region (e.g., \citealt{peterson93, richards2002}). TDEs generally lack \ion{Mg}{2} emission, including ASASSN-14li, iPTF15af, iPTF16fnl, and possibly ASASSN-18jd. \ion{Mg}{2} was detected in absorption in ZTF19abzrhgq, but \citet{hung2020} attributed this feature to the host ISM and circumnuclear gas instead of stellar debris associated with the TDE. In ASASSN-14ko, \ion{Mg}{2} is not strongly detected at $-4$, $+0$, and $+7$ days but it is clearly detected at $-36$ days. This trend is another instance of the UV spectra looking more TDE-like during the flares and more AGN-like during quiescence. Since \ion{Mg}{2} originates in partially ionized regions, the development of a luminous hard continuum should suppress \ion{Mg}{2} emission. The \ion{Mg}{2} line is always blueshifted from its expected rest frame wavelength, likely as a result of host galaxy dynamics as was also discussed in \citet{tucker2020}. \ion{C}{3}], which is used to constrain the density of the broad-line region \citep{ferland1992, peterson1997}, is detected at all times in ASASSN-14ko, but not in any other TDE spectra. It was previously observed that the UV spectra of ASASSN-14li and iPTF16fnl are similar to N-rich QSOs \citep{brown2018}.   

Figure \ref{fig:BH_scales_R_V} tries to place the various radii and velocities in context given the rough estimates of the SMBH mass from \citet{payne2021a}. As discussed there, the Roche limits for main sequence stars, here illustrated by an $0.3M_\odot$ dwarf and the Sun, are very close to the Schwarzschild radius.  Giants, illustrated using a star with a radius of $10 R_\odot$, have Roche limits well outside the horizon.  The apparent orbital period corresponds to a semi-major axis roughly ten times larger than the Roche limit of such a star, which would be consistent with a star on an elliptical orbit being periodically stripped at pericenter. Note that a much larger star ($\sim 100 R_\odot$) would be inconsistent with this picture because its Roche limit would be very similar to the semimajor axis.  The range of black body radii shown in Figure \ref{fig:bbody_phasefold} corresponds to a scale very similar to the Roche limit of the $10R_\odot$ star. 

If we interpret the expansion of the black body radius from $10^{14.2}$ to $10^{14.9}$~cm over 8 days as a physical motion, it implies a velocity of $34000$~km/s.  As a circular velocity, this corresponds to a radius modestly larger than the black body radius and the giant Roche limit and modestly smaller than the semi-major axis.  The typical absorption and emission line velocities of $\sim 5000$~km/s, again interpreted as a circular velocity, correspond to radii an order of magnitude larger than the semi-major axis.  The orbital time scales at these radii are decades, so the changes in the emission and absorption lines during a flare cannot be driven by physical  motions associated with the flare -- they must be due to ionization changes driven by changes in the luminosity and spectrum during the flare.  The light travel times to these radii are several days to a week, so it should be possible to use intensive UV spectral monitoring during a flare to make a reverberation mapping measurement of the characteristic distance of the emission lines (absorption lines, by definition, have zero lag).

The six flares with X-ray observations are all characterized by an X-ray luminosity dimming and spectral hardening a few days before the optical peak. The column density does not change significantly, which rules out the possibility that the change in X-ray luminosity and spectrum is driven by a change in the absorption. ASASSN-14ko's X-ray/UV/optical behavior is similar to ASASSN-18el (AT2018zf; \citealt{nicholls2018}) which underwent a drop in X-ray luminosity by several orders of magnitude in conjunction with the UV/optical brightening \citep{trakhtenbrot201918el, ricci2020, ricci2021}. Different scenarios have been advanced to explain ASASSN-18el's behavior. It could be driven by an instability within the AGN accretion disk that causes a change in the accretion rate \citep{trakhtenbrot201918el} or an inversion of magnetic flux in a magnetically arrested disk \citep{scepi2021, laha2022}. Alternatively, ASASSN-18el was a TDE that destroyed the X-ray corona and inner accretion disk \citep{ricci2020,ricci2021} leading to the decreased X-ray luminosity and increased UV/optical emission. The X-ray luminosity then recovers as the X-ray corona reforms. However, for ASASSN-18el the X-ray spectrum became softer rather than harder as it faded, and the fading occurred long after the optical peak and lasted for longer. 

\section{Conclusion} \label{conclusion}
ASASSN-14ko is a predictable nuclear transient whose flares are well-characterized by a timing model, although the individual flares are not identical. They do not, however, show an even/odd dichotomy which we might expect from a star disrupting an accretion disk. Using ASAS-SN, \textit{Chandra}, \textit{HST}/STIS, \textit{NICER}, \textit{Swift} X-ray/UVOT, and \textit{TESS}, we presented the photometric and spectroscopic X-ray/UV evolution and photometric optical evolution. Our findings can be summarized as follows: 

\begin{itemize}
    \item We refit the timing model first used in \citet{payne2021a} to include the recent flare events and we find period $P_0 = 115.2^{+1.3}_{-1.2}$ days and period derivative $\dot{P} = -0.0026 \pm 0.0006 $.
    \item The UV/optical light curves always brighten to a single large-amplitude peak, but the peak luminosity varies between peaks. The UV luminosity peaks show larger differences than the optical luminosity peaks. This is also reflected in the blackbody luminosities. 
    \item The two \textit{TESS} light curves from Sectors 3-5 and 31-33 show that the optical flares are not truly identical. The earlier flare began to rise earlier but more slowly to a fainter peak and then declined slower. 
    \item The X-ray luminosity consistently decreases during the UV/optical rise, but the depth of the minimum varies. Across all flares, the X-ray emission is consistently harder when fainter and softer when brighter. There seems to be no associated change in the absorption.
    \item The \textit{Chandra} data showed that the northern nucleus of the host galaxy brightened during the flare, indicating that this nucleus is the source of ASASSN-14ko's X-ray emission. For both epochs, the absorbed power-law \texttt{xspec} model $\textrm{tbabs} \times \textrm{zashift} \times \textrm{powerlaw}$ with a photon index of $1.34\pm0.11$ at $+4$ days and $1.32\pm0.09$ at $+26$ days provided the best fit for ASASSN-14ko.
    \item Both the UV continuum and UV spectral lines changed rapidly during the December 2020 flare as revealed by \textit{HST}/STIS. The UV spectral lines that evolved the most dramatically were Ly$\alpha$, \ion{N}{5}, \ion{Si}{4}, and \ion{C}{4} which showed broad absorption features at $\sim 5000$ km/s four days before the optical peak which vanished and were replaced by broad emission lines by seven days after peak. Overall, the UV spectra show some similarities to other UV TDE spectra during outburst but they are more similar to AGN spectra in quiescence. 
\end{itemize}

We are continuing to monitor the flares and have proposed to obtain more complete UV spectra phase coverage. We have an extensive body of optical spectra and these will be analyzed in Payne et al. (in prep).  

\vspace{5mm}

{\bf Software:} astropy \citep{astropy2018}, ftools \citep{blackburn95}, HEAsoft \citep{heasarc2014}, IRAF \citep{tody1986, tody1993}, numpy \citep{harris2020}, matplotlib \citep{hunter2007} 

\section{Acknowledgements}

We thank Rick Pogge for valuable discussions. We thank Las Cumbres Observatory and its staff for their continued support of ASAS-SN. ASAS-SN is funded in part by the Gordon and Betty Moore Foundation through grants GBMF5490 and GBMF10501 to the Ohio State University, and also funded in part by the Alfred P. Sloan Foundation grant G-2021-14192. 
A.V.P. acknowledges support from the NASA Graduate Fellowship through grant 80NSSC19K1679. B.J.S., C.S.K., and K.Z.S. are supported by NSF grant AST- 1907570. B.J.S. is also supported by NSF grants AST-1920392, AST-1911074, and AST-1911074. C.S.K. and K.Z.S. are supported by NSF grant AST-181440. Support for T.W.-S.H. was provided by NASA through the NASA Hubble Fellowship grant HST-HF2-51458.001-A awarded by the Space Telescope Science Institute, which is operated by the Association of Universities for Research in Astronomy, Inc., for NASA, under contract NAS5-26555. Parts of this research were supported by the Australian Research Council Centre of Excellence for All Sky Astrophysics in 3 Dimensions (ASTRO 3D), through project number CE170100013. T.A.T. is supported in part by Scialog Scholar grant 24215 from the Research Corporation.  J.T.H. is supported by NASA award 80NSSC21K0136.

\bibliographystyle{aasjournal}
\bibliography{references}

\appendix 

\begin{figure}
    \centering
    \includegraphics[width=0.95\linewidth]{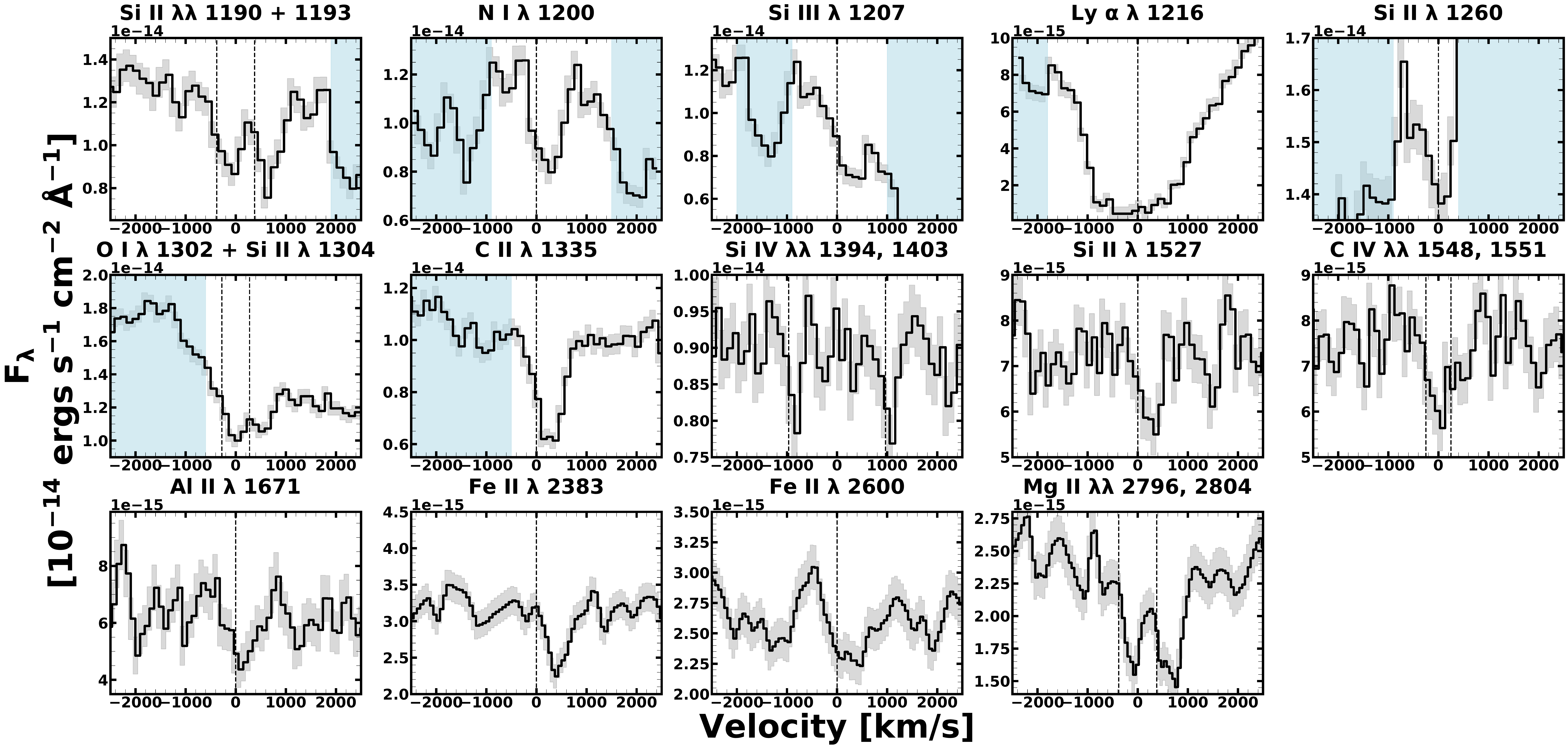}
    \caption{Low- and high-ionization absorption lines attributed to the Milky Way ISM shown for the quiescence spectrum observed on 2021-03-06. The velocity scale is at rest-frame wavelength at $z = 0$ for each line, or in the case of blended or doublet lines, an average of the two rest-frame wavelengths. Each velocity is indicated by the vertical dashed lines. The shaded blue regions are areas of the spectrum affected by neighboring absorption and emission lines. The gray shaded regions show the spectrum uncertainty. }
    \label{fig:HST_MWlines}
\end{figure}

In addition to emission and absorption features detected at the redshift of the host, the UV \textit{HST}/STIS spectra exhibit low- and high-ionization absorption lines at $z = 0$. We attribute these features shown in Figure \ref{fig:HST_MWlines} as originating from the Milky Way interstellar medium (ISM). These transitions consist of low-ionization elements (ionization energy $<13.6$ eV) \ion{N}{1}, \ion{Si}{2}, Ly$\alpha$, \ion{C}{2}, \ion{Fe}{2}, \ion{Mg}{2}, \ion{Al}{2}, and \ion{O}{1}, in addition to high-ionization elements (ionization energy $>13.6$ eV) \ion{S}{3}, \ion{Si}{4} and \ion{C}{4}. Numerous lines are affected by neighboring absorption and emission lines, but all absorption lines from the Galactic ISM are less than $500~\mathrm{km}~\mathrm{s}^{-1}$.  

\end{document}